\documentclass[12pt]{article}
\pdfoutput=1
\usepackage{jheppub}
\usepackage{epsfig,graphicx,xcolor,amsbsy,amssymb,latexsym,amsfonts,amsmath,setspace}

\usepackage[normalem]{ulem}

\usepackage{pstricks}
\usepackage{color}
\usepackage{soul}

\usepackage{wrapfig}
\usepackage{tikz}

\usepackage{pst-node}
\usepackage{tikz-cd}
\usepackage{subfig}

\usepackage{graphicx}

\usepackage{xcolor}

\usepackage[enableskew]{youngtab}
  
\usepackage{etex}
\usepackage{relsize}
\usepackage{amsfonts}
\usepackage[T1]{fontenc}
\usepackage{hyperref}

\usepackage{blindtext}

\def\bea{\begin{eqnarray}}
\def\eea{\end{eqnarray}}

\usepackage{appendix}

\usetikzlibrary{decorations.markings,arrows,backgrounds,patterns,decorations.pathreplacing}

\preprint{LCTP-24-24}

\title{A Four-dimensional Gauge Theory Perspective on Quantum K-theory}

\author{M. Nouman Muteeb}

\author[a]{and Leopoldo A. Pando Zayas}

\emailAdd{nouman01uet@gmail.com, lpandoz@umich.edu}

\affiliation[a]{Leinweber Center for Theoretical Physics, 
University of Michigan, Ann Arbor, MI 48109, USA}

\affiliation[a]{The Abdus Salam International Centre for Theoretical Physics, 34014 Trieste, Italy}

\abstract{ The two-dimensional gauged linear sigma model has provided a physical model for the quantum cohomology of a K\"ahler manifold, $X$. A three-dimensional version of such construction has recently been shown to shed light on models of quantum K-theory of $X$. We consider an $\mathcal{N}=1$ four-dimensional version consisting of a $U(1)$ vector multiplet and chiral multiplets, generalizing the two-dimensional $\mathcal{N}=(2,2)$ setup. We compute the four-dimensional partition function on $D^2\times \mathbb{T}^2$ and demonstrates that it satisfies a difference equation which reduces to the deformed quantum K-theoretic one in the appropriate limit. We also demonstrate, though indirectly, that 4d invariants reduce to 3d quantum K-theory invariants in the same limit. }

\date{\today}

\begin{document}

\maketitle

\section{Introduction}
The two-dimensional gauged linear sigma model (GLSM) provides a physical model of quantum cohomology on a K\"ahler manifold $X$.  From the physics point of view, the quantum product of certain chiral operators defines a deformation of the classical intersection ring, leading to impressive insights into enumerative geometry. Recently, the 2d GLSM was uplifted to a 3d ${\cal N}=2$ gauge theory defined on the twisted product of a circle and a disk, $D^2\times_q S^1$, where $q$ is the weight of the twisting. The theory was shown to provide a physical model for Givental's  permutation-equivariant quantum K-theory. Jockers and Mayr  were able to recover the 2d GLSM description of quantum cohomology in the small-radius limit, $q\to 1$ 
 \cite{Jockers:2018sfl}. With this manuscript, we initiate a series exploring the implications of naturally uplifting the 2d GLSM to a 4d gauge theory. The main result reported here is to show that the partition function of the 4d gauge theory on the twisted product $D^2\times_{(q,p)} \mathbb{T}^2$ provides an explicit construction of a one-parameter refinement of Givental's  permutation-equivariant quantum K-theory.

The central paradigm in every situation is the connection between the counting of supersymmetric BPS states and geometric information of the K\"ahler manifold, $X$, which is the target of the theory. An important element of this paradigm is the independence of the BPS counting on the energy scale of the theory. A simple physical ultraviolet (UV) model for quantum cohomology is the GLSM \cite{Witten:1993yc,Morrison:1995yh} --   $\mathcal{N}=(2,2)$ supersymmetric 2d gauge theory, which flows in certain phases to the non-linear sigma model in the infrared (IR). In the 3d case, Jockers and Mayr considered an $\mathcal{N}=2$ supersymmetric lift of the 2d GLSM and found the associated ring structure \cite{Jockers:2018sfl}. It was further shown that this 3d UV gauge theory also computes a topological theory in the IR. The theory in question replaces the side of quantum cohomology in the 2d correspondence. The authors showed that the IR theory is the permutation-equivariant quantum K-theory constructed by Givental \cite{Givental2015PermutationequivariantQK1,Givental2015PermutationequivariantQK2,Givental2015PermutationequivariantQK3,Givental2015PermutationequivariantQK4,Givental2015PermutationequivariantQK5,Givental2015PermutationequivariantQK6,Givental2015PermutationequivariantQK7,Givental2015PermutationequivariantQK8,Givental2015ExplicitRI}. For other interesting and related works see for example \cite{Jockers:2012dk,Jockers:2021omw,Jockers:2019lwe}.

In this note, we are motivated to explore the naturally arising 4d framework. Namely, we consider a 4d $\mathcal{N}=1$ gauge theory and study how it deforms some of the quantum K-theory structures obtained in the 3d context by Jockers and Mayr. The Quantum K-theory of a K\"ahler manifold is a collection of integers that is usually assembled into a generating series $J(Q,q,t)$ that satisfies a system of linear differential equations with respect to $t$ and $q$-difference equations with respect to $Q$. Under certain conditions, the full theory can be reconstructed from its small $J$-function $J(Q,q,0)$; reproducing this small $J$-function from the 4d perspective is the subject of this manuscript.

After this brief introduction, we provide a summary of equivariant quantum K-theory in Section \ref{Sec:PE-QK}. In Section \ref{Sec:Localization} we describe the computation of a $U(1)$ 4d gauge theory partition function on $D^2\times_{(q,p)} \mathbb{T}^2$. The matter content of this theory is the 4d $\mathcal{N}=1$ uplift of the 2d GLSM with the quintic Calabi-Yau 3-fold as target space.  In Section \ref{Sec:Diff-Eqn} we propose a 4d difference operator which annihilates the $J$-function corresponding to the 4d gauge theory. The $ q\to 0$ limit of the 4d partition function and 4d difference operator are discussed in Section \ref{Sec:qto0}. We show that in this limit one gets a certain deformation of the 3d partition function and 3d difference equation discussed in \cite{Jockers:2018sfl}.  Section \ref{Sec:4dgenQK} presents some details regarding how the 4d big $J$-function transformed by generalized (4d) Givental's transformations reproduces the correct 3d small $J$-function $J^{3d}(0)$ of \cite{Jockers:2018sfl} in the appropriate $q\to 0$ limit. Recall that $J^{3d}(0)$ along with the appropriate Givental transformation is what we need to create all 3d invariants. This implies, indirectly, that the 4d quantum K-theory invariants reduce to the correct 3d quantum K-theory invariants.


\section{Permutation Equivariant Quantum K-theory: A Summary}\label{Sec:PE-QK}
The physical origin of Gromov-Witten theory lies in string theory. There, it corresponds to scattering amplitudes in a topological version of string theory, called A-model topological string theory.
We give the definition following \cite{Givental2011TheHT,Cox:2000vi} over the field of complex numbers $\mathbb{C}$. Let us denote by $X_{g,n,d}$ the moduli space of stable maps for manifold {\color{blue}\sout{$Y$}} $X$. The moduli spaces $X_{g,n,d}$ parametrize the data given by $(C,x_1,...,x_n,f)$, where $C$ represents genus $g$ Riemann surface with $n$ marked points, $f:C\to X$ denotes a holomorphic map of degree $d\in H_2(Y,\mathbb{Z})$. One can define natural maps
\bea 
ev_i:X_{g,n,d}\to X,\quad i=1,...,n
\eea
There are also natural line bundles
\bea 
L_i\to X_{g,n,d}\quad i=1,...,n
\eea
dubbed universal cotangent line bundles. The cotangent line to $C$ at the point $x_i$ forms the fiber of $L$ over $(C,x_1,...,x_n,f)$. Consider the generators $a_1,...,a_n\in K^0(X,\mathbb{C})$. K-theoretic Gromov-Witten invariants are defined as the holomorphic Euler characteristics of the following sheaves over $X_{g,n,d}$
\bea 
ev_1^*(a_1)L_1^{k_1}...ev_n^*(a_n)L_n^{k_n}\otimes \mathcal{O}^{vir}.
\eea
In the correlator notation we have
\bea 
<a_1L_1^{k_1}...a_nL_n^{k_n}>_{g,n,d} := \chi(X_{g,n,d};ev_1^*(a_1)L_1^{k_1}...ev_n^*(a_n)L_n^{k_n}\otimes \mathcal{O}^{vir}).
\eea
The sheaf $\mathcal{O}^{vir}$ called the virtual structure sheaf plays a role in K-theory analogous to virtual fundamental class $[X_{g,n,d}]$ in cohomological Gromov-Witten theory.

The genus zero quantum K-theory invariants of a K\"ahler manifold $X$ are defined as the holomorphic Euler characteristics of a certain moduli space of stable maps $\bar{\mathcal{M}}_{0,m}(X,\beta)$
of class $\beta\in H_2(X,\mathbb{Z})$
\bea 
<t_1(q),,,t_m(q)>_{0,m,\beta}=\chi_{\bar{\mathcal{M}}_{0,m}(X,\beta)}(ev_1^*t_1(L_1)ev_2^*t_2(L_2)...ev_m^*t_n(L_m)\otimes\mathcal{O}^{vir})
\eea
where $t_i(q)\in K(Y)[q,q^{-1}]$ and depend on the cotangent line bundles over $(C,x_1,...,x_m,f)\in \bar{\mathcal{M}}_{0,m}(X,\beta)$  in the moduli space of $n$-pointed genus-$g$ stable maps of class $\beta$ with marked point $x_i$ and $ev_i:\bar{\mathcal{M}}_{0,m}(X,\beta) \to X$ denotes the evaluation map at the point $x_i$. The virtual structure sheaf $\mathcal{O}^{vir}$ is the K-theory equivalent of the virtual fundamental class of quantum cohomology.

Quantum K-theory is the counterpart of Gromov-Witten theory where holomorphic Euler characteristics of  vector bundles over moduli spaces of holomorphic curves replaces the
 cohomological intersection numbers on such moduli spaces. The theory is in some ways parallel to the cohomological one and brings up some new features. It turns out that in order to define the theory consistently it is necessary to consider the action of the group of permutations of the marked points. It implies the decomposition of sheaf cohomology of vector bundles over moduli spaces according to representations of the permutation groups. One defines `permutation-equivariant' K-theoretic GW-invariants as multiplicities of such representations. The computation of GW-invariants using tori actions and fixed point localization in  K-theory necessitates such information. In Givental's construction of permutation equivariant Quantum K-theory the q-analogues of toric hyper-geometric functions of cohomological mirror theory only arise in K-theoretic version of mirror theory if the action of permutation groups is incorporated. Interestingly, K-theoretic Gromov-Witten invariants are integers. Contrast this with cohomological Gromov-Witten invariants which are rational numbers.
 
Similarly, equivariant Quantum K-theory invariants are defined by
\bea 
<t_1(q),,,t_{m-n}(q);t(q),...,t(q)>^{S_n}_{0,m,\beta}=\sum_{\nu\in Irrep(S_n)}\chi_{\beta}^{S_n,\nu}(t_1(q),,,t_{m-n}(q);t(q))\nu,\nonumber\\
\eea
where $\chi^{S_n}$ is the permutation-equivariant Euler characteristic. The Quantum K-theory correlator can also be described as a K-theoretic pushforward map \cite{Givental2015PermutationequivariantQK1}
\bea
&&(t(q),...,t(q))^{S_n}_{g,n,d}=\nonumber\\&&\big(\pi:(X_{g,n,d}\times Y)/ S_n\to Y \big)_*\Big( \mathcal{O}^{virt}_{g,n,d}\otimes\prod_{i=1}^n\big( \sum_{m\in\mathbb{Z}}ev_i^*(t_m)L^m_i\big) \Big)
\eea
where $t_m$ are coefficients in the Laurent expansion  $t(q)=\sum_{m\mathbb{Z}}t_m q^m$.\\
The 4d point of view leads to considering the more general Laurent expansion $t(q_1,q_2)=\sum_{m,n\in \mathbb{Z}}t_{m,n}q_1^nq_2^m$. It is natural to assume that the above correlator will be modified to
\bea
&&(t(q_1,q_2),...,t(q_1,q_2))^{S_n}_{g,n,d}=\nonumber\\&&\big(\pi:(X_{g,n,d}\times Y)/ S_n\to Y \big)_*\Big( \mathcal{O}^{virt}_{g,n,d}\otimes\prod_{i=1}^n\big( \sum_{m\in\mathbb{Z}}ev_i^*(t_{m,n})L^m_iP_i^n\big) \Big).
\eea
In this manuscript we treat the above expression in terms of formal variables and a precise mathematical interpretation will be discussed elsewhere.  



\section{4d partition function of $D^2\times \mathbb{T}^2$}\label{Sec:Localization}

The partition function of 4d theories on $D^2\times \mathbb{T}^2$ has been discussed previously in \cite{Nishioka:2014zpa,Longhi:2019hdh}  and for $S^2\times \mathbb{T}^2$ in \cite{Closset:2013sxa,Honda:2015yha}. The background metric can be given in real coordinates as

\bea\label{Eq:T2D2}
ds^2= (dx+{\rm Re}(\tau)dy)^2+{\rm Im}(\tau)^2 dy^2+d\vartheta^2+\sin^2\vartheta(d\phi+\alpha dx+\beta dy)^2
\eea
where $\tau\in\mathbb{H}^+$ is the modular parameter of $\mathbb{T}^2$, $\alpha,\beta \in \mathbb{R}$, $x \sim x+1, y\sim y+1$, $\vartheta\in[0,\pi/2]$ and $\phi\sim \phi+2\pi$. Following the paradigm of rigid supersymmetry introduced by Festuccia and Seiberg \cite{Festuccia:2011ws}, we start by considering supersymmetry on this background as defined by the following Killing spinor equations origininating in the new minimal off-shell supergravity:
\bea 
(\nabla_{\mu}-iA_{\mu}+i V_{\mu}+i\sigma_{\mu\nu}V^{\nu})\zeta&=&0,\nonumber\\
(\nabla_{\mu}+iA_{\mu}-i V_{\mu}+i\tilde{\sigma}_{\mu\nu}V^{\nu})\tilde{\zeta}&=&0,\nonumber\\
\eea
where $V_{\mu}$ is the background field satisfying $\nabla_{\mu}V^{\mu}=0$ and $A_{\mu}$ denotes the background Abelian connection of R-symmetry. It is convenient to highlight the complex structure by considering the manifold as defined by the following global identifications $(\omega, z)\sim (\omega+2\pi, e^{2\pi i \alpha} z)\sim (\omega+2\pi \tau, e^{2\pi i \beta} z)$ and the following metric:
\bea\label{Eq:ComplexCoord}
ds^2=d\omega d\bar{\omega}+\frac{4 dz d\bar{z}}{(1+|z|^2)^2}.
\eea
The complex moduli structure is then given by $\tau$ and $\sigma = \alpha \tau -\beta$. 
The above metric requires the following values of background fields
\bea 
V=0,\quad A=-\frac{i}{(1+|z|^2)}(\bar{z}dz-zd\bar{z}) -\frac{i d\text{log s}}{2},
\eea
where $s$ is the nowhere vanishing global section of R-symmetry line bundle tensored with the canonical line bundle $\mathcal{R}^2\otimes\mathcal{K}$ and for our background we make the choice of constant $|s|$ to keep $A$ real. The canonical line bundle $\mathcal{K}$ for our background is trivial and we do not need to quantize R-charges.\\
The reduction of the geometry $D^2\times \mathbb{T}^2$ to the geometry $D^2\times S^1$ is more manifest in real coordinates introduced in \eqref{Eq:T2D2}:
\bea
ds^2&=& dx^2 +d\vartheta^2+\sin^2\vartheta (d\phi+\alpha dx)^2 \nonumber \\
&+&\left({\rm Re}(\tau)^2+{\rm Im}(\tau)^2 + \beta^2\sin^2\vartheta\right)dy^2+ 2 dy\left[{\rm Re}(\tau)dx+\beta\sin^2\vartheta(d\phi+\alpha dx)\right]. \nonumber 
\eea
Following the Kaluza-Klein reduction paradigm, the first line is the 3d metric used in \cite{Jockers:2018sfl}, the second line can be interpreted as a scalar field and a gauge field from the 3d point of view. It is worth highlighting that the role of the twist $\alpha$ is clear from the construction of Jockers and Mayr, it corresponds to the parameter $q$ in their notation of $D^2\times_q S^1$ \cite{Jockers:2018sfl}. Here we have a new parameter, $\beta$, leading to the generalization we seek. It is, however, worth recalling that the complex structure moduli are $\tau$ and $\sigma=\alpha \tau -\beta$ and the partition function will depend only on these combinations although it might make sense intuitively to think of the twisting as  $D^2\times_{(\alpha, \beta)} \mathbb{T}^2$. Moreover, in regard to the line bundles, note that in real coordinates the background gauge field takes the form
\begin{equation}
    A=\frac{1}{2}(1-\cos\vartheta)(d\phi+\alpha\, dx +\beta \,dy).
\end{equation}
This expression indicates that the refinement provided by $\alpha$ and $\beta$ is equivalent to turning on a flat connection in the $d\phi$ direction of the disk, essentially an equivariant deformation of the chemical potential, $\sigma$, for the angular momentum on the disk \cite{Longhi:2019hdh}.
\subsection{One-loop determinant of chiral multiplet}

In 2d case it was shown that  the K\"aler potential of the A-model contains information about the genus zero Gromov-Witten invariants of $X$ \cite{Jockers:2012dk,Gomis:2012wy}. We are interested in exploring similar structures on X in the case where we take all the multiplets of 2d GLSM and promote them to 4d $\mathcal{N}=1$ multiplets, leading to the 4d $\mathcal{N}=1$ version of 2d $\mathcal{N}=(2,2)$ gauge linear sigma models with target space $X$. The 4d equivalent of 2d GLSM is put on the twisted $D^2\times \mathbb{T}^2$ background. The quantities of interest now are the quantum products of certain 4d gauge theory operators associated with the K-theory group $K(X)$ on the K\"ahler manifold $X$. We proceed by determining the partition function. 

The one-loop determinants of 4d $\mathcal{N}=1$ multiplets with supersymmetric Dirichlet and Robin boundary conditions are computed in, for example,  \cite{Longhi:2019hdh}. The 4d gauge theory of interest is a $U(1)$ gauge theory, so the one-loop determinant of only chiral multiplets are relevant,

\bea
Z_{\rm 1-loop}^{{\rm chi}(D)}={\rm det}_{\mathcal{R}}[\frac{e^{-i\frac{\pi}{3}P_3(\sigma(1-r/2)-\Phi_0)}}{\Gamma(\sigma(1-r/2)-\Phi_0;\sigma,\tau)}],
\eea
\bea
Z_{\rm 1-loop}^{{\rm chi}(R)}={\rm det}_{\mathcal{R}}[e^{i\frac{\pi}{3}P_3(\sigma(r/2)+\Phi_0)}\Gamma(\sigma(1-r/2)-\Phi_0;\sigma,\tau)],
\eea
where $D$ denotes Dirichlet boundary conditions, $R$ denotes Robin boundary conditions, $\Gamma(\sigma(1-r/2)-\Phi_0;\sigma,\tau)$ is the elliptic gamma function and $\Phi_0$ is the zero mode of the gauge connection along the Killing vector field on the torus.

\subsection{4d lift of the Sigma Model for the Quintic Calabi Yau 3-fold}

The quintic Calabi-Yau 3-fold $Y$ is defined as a degree 5 hypersurface in the projective space $\mathbb{P}^4$. The classical K-theory ring $K(Y)$ has four generators $\Phi_k=(1-P)^k$, $k=0,1,2,3$, where $P$ is defined as the restriction of the tautological line bundle $\mathcal{O}(-1)$ of $\mathbb{P}^4$ to the hypersurface $Y$. The intersection pairing of the generators is given by the following $4\times4$ matrix
\bea
(\Phi_k,\Phi_l)=\int_Y \text{td}(Y)\text{ch}(\Phi_k\otimes\Phi_l)=\begin{bmatrix}
0 & 5 & -5 & 5 \\
5 & -5 & 5 & 0 \\
-5 & 5 & 0 & 0 \\
5 & 0 & 0 & 0
\end{bmatrix}.
\eea

The 4d gauge theory with the quintic $Y$ as target  is a $U(1)$ Abelian gauge theory with six $\mathcal{N}=1$ chiral multiplets. Five of the multiplets have $U(1)$ gauge charge $1$ and R-charge $0$. The sixth chiral multiplet has gauge charge $-5$ and R-charge $2$. The one-loop determinant for this gauge theory on $D^2\times \mathbb{T}^2$ is given by

\bea
Z_{\rm 1-loop}^{\rm Quintic}=e^{-\frac{i\pi}{3}(P_3(5u)-5P_3(u))}\frac{\Gamma(u;\sigma,\tau)^5}{\Gamma(5u;\sigma,\tau)},
\eea
where 
\bea
P_3(X)&=&\frac{X^3}{\tau\sigma}-\frac{3(1+\tau+\sigma)X^2}{2\tau\sigma}+\frac{(1+\tau^2+\sigma^2+3(\tau+\sigma+\tau\sigma))X}{2\tau\sigma}-\frac{(1+\tau+\sigma)(\tau+\sigma+\sigma\tau)}{4\tau\sigma}\nonumber\\
&-&\frac{1-\tau^2+\tau^4}{24\sigma(\tau+\tau^2)}.
\eea
Using the following definitions 
\bea
\Gamma(x;p,q)=\frac{(pqx^{-1};p,q)_{\infty}}{(x;p,q)_{\infty}},
\eea
\bea
(x;p,q)_{\infty}=\prod_{j=0}\prod_{k=0}(1-x p^j q^k),
\eea
for $|p|<1,|q|<1$, we get the expression for one-loop determinant in the following form
\bea
Z_{\rm 1-loop}^{\rm Quintic}=e^{-\frac{i\pi}{3}(P_3(5u)-5P_3(u))}\prod_{j=0}^{\infty}\prod_{k=0}^{\infty}\frac{(1-\lambda_-s^5\tilde{P}^jq^k)}{(1-\lambda_+s\tilde{P}^jq^k)^5}\prod_{j=0}^{\infty}\prod_{k=0}^{\infty}\frac{(1-\lambda_+^{-1}s^{-1}\tilde{P}^{j+1}q^{k+1})^5}{(1-\lambda_-^{-1}s^{-5}\tilde{P}^{j+1}q^{k+1})},\nonumber\\
\eea
where $s=e^{2\pi i u}$, $\tilde{P}=e^{2\pi i \sigma}$ and $q=e^{2\pi i \tau}$; $\tau$ is the modular parameter of the torus. The partition function is given as an integral over $u$ variable describing the gauge field
\bea 
Z^{\rm Quintic}=\int \frac{du}{2\pi i}Z_{\rm 1-loop}^{\rm Quintic}.
\eea
The poles of the integrand in the $u$-plane are located at
\bea
1-\lambda_+s\tilde{P}^jq^k=0&:&\quad  s=\lambda_+^{-1}\tilde{P}^{-j_1}q^{-k_1}\tilde{P}^{\epsilon},\nonumber\\
1-\lambda_-^{-1}s^{-5}\tilde{P}^{j+1}q^{k+1}=0&:&\quad s=\lambda_-^{-1/5}\tilde{P}^{(-j_1-1)/5}q^{(-k_1-1)/5}\tilde{P}^{-\epsilon/5}.
\eea
According to the Jeffrey-Kirwan residue prescription \cite{Benini:2013xpa}, we can choose the poles at $s=\lambda_+^{-1}\tilde{P}^{-j_1}q^{-k_1}\tilde{P}^{\epsilon}$ or we can choose the poles at $s=\lambda_-^{-1/5}\tilde{P}^{(-j_1-1)/5}q^{(-k_1-1)/5}\tilde{P}^{-\epsilon/5}$. At the end, the two prescriptions should be equivalent. Note that we have introduced a new variable $\epsilon$ and have redefined the location of poles.\\
Our choice is the first set of poles. After substitution we get the following expression for the partition function
\bea
Z^{\rm Quintic}&=&\int\frac{d\epsilon}{2\pi i}\sum_{j_1=0}^{\infty}\sum_{k_1=0}^{\infty}Q_1^{j_1}Q_2^{k_1}e^{-i\frac{\pi}{3}(120a(-\sigma j_1+\epsilon\sigma-\tau k_1)^3+20b(-\sigma j_1+\epsilon\sigma-\tau k_1)^2)}\nonumber\\&\times&e^{\frac{- i \pi }{3}(\frac{4(1+\tau+\sigma)(\tau+\sigma+\tau\sigma)}{4\tau\sigma}+\frac{1-\tau^2+\tau^4}{6\sigma(\tau+\tau^2)})}
\nonumber\\&\times&
\prod_{j=0}^{\infty}\prod_{k=0}^{\infty}\frac{1-\tilde{P}^{j-5j_1+5\epsilon}q^{k-5k_1}}{(1-\tilde{P}^{j-j_1+\epsilon}q^{k-k_1})^5}\prod_{j=0}^{\infty}\prod_{k=0}^{\infty}\frac{(1-\tilde{P}^{j+j_1+1-\epsilon}q^{k+k_1+1})^5}{1-\tilde{P}^{j+5j_1+1-5\epsilon}q^{k+5k_1+1}},
\eea
where $a=\frac{1}{\tau\sigma}$, $b=-\frac{1+\tau+\sigma}{2\tau\sigma}$.  Here we have made used of $Q_1$ and $Q_2$, the so-called  Novikov variables, they elucidate the contribution of various curves in $X$ by their degrees. Note that there is an overall factor $e^{\frac{- i \pi }{3}(\frac{4(1+\tau+\sigma)(\tau+\sigma+\tau\sigma)}{4\tau\sigma}+\frac{1-\tau^2+\tau^4}{6\sigma(\tau+\tau^2)})}$. We can normalize the partition function by this factor (recall that $q=e^{2\pi i \tau}$). So we can consider the normalized partition function again denoted by $Z^{\rm Quintic}$,
\bea\label{eq:quinticnorma}
Z^{\rm Quintic}&=&\int\frac{d\epsilon}{2\pi i}\sum_{j_1=0}^{\infty}\sum_{k_1=0}^{\infty}Q_1^{j_1}Q_2^{k_1}e^{-i\frac{\pi}{3}(120a(-\sigma j_1+\epsilon\sigma-\tau k_1)^3+20b(-\sigma j_1+\epsilon\sigma-\tau k_1)^2)}
\nonumber\\&\times&
\prod_{j=0}^{\infty}\prod_{k=0}^{\infty}\frac{1-\tilde{P}^{j-5j_1+5\epsilon}q^{k-5k_1}}{(1-\tilde{P}^{j-j_1+\epsilon}q^{k-k_1})^5}\prod_{j=0}^{\infty}\prod_{k=0}^{\infty}\frac{(1-\tilde{P}^{j+j_1+1-\epsilon}q^{k+k_1+1})^5}{1-\tilde{P}^{j+5j_1+1-5\epsilon}q^{k+5k_1+1}}.
\eea
In the literature on K-theoretic Gromov-Witten invariants of a manifold $X$, the integrand of the partition function is called the J-function, therefore, 
\bea
J^{4d}&=&\sum_{j_1=0}^{\infty}\sum_{k_1=0}^{\infty}Q_1^{j_1}Q_2^{k_1}e^{-i\frac{\pi}{3}(120a(-\sigma j_1+\epsilon\sigma-\tau k_1)^3+20b(-\sigma j_1+\epsilon\sigma-\tau k_1)^2)}\nonumber\\&\times&
\prod_{j=0}^{\infty}\prod_{k=0}^{\infty}\frac{1-\tilde{P}^{j-5j_1+5\epsilon}q^{k-5k_1}}{(1-\tilde{P}^{j-j_1+\epsilon}q^{k-k_1})^5}\prod_{j=0}^{\infty}\prod_{k=0}^{\infty}\frac{(1-\tilde{P}^{j+j_1+1-\epsilon}q^{k+k_1+1})^5}{1-\tilde{P}^{j+5j_1+1-5\epsilon}q^{k+5k_1+1}}.
\eea
$J-$functions are used to represent  generating function of K-theoretic Gromov-Witten invariants. More precisely, the above $J$-function is called the big $J$-function. Schematically it is given by
\bea 
J(t) := (1-q)+t(q)+\sum_{a}\Phi^a\sum_{n,d}\frac{Q^d}{n!}<\frac{\Phi_a}{1-qL},t(L),...,t(L)>^X_{d,0,n+1}
\eea
The term $1-q$ is called the dilaton shift, $t(q)$ is called the $input$, and the remaining terms are the source of Gromov-Witten invariants. The pair $\Phi^a$ and $\Phi_a$ form Poincare dual basis of $K^0(X)$. The term $(1-q)+t(q)$ lies in the space of Laurent polynomials denoted as $\mathcal{K}_+$ and the remaining terms lie in a space of functions which are regular at $q=0$ and vanish at $q=\infty$, denoted as $\mathcal{K}_-$. In the loop space formalism, the loop space is defined by
\bea 
\mathcal{K} := K^0(X)\otimes \mathbb{C}[q,q^{-1}]\otimes \mathbb{C}[[Q]]
\eea
Using the K-theoretic intersection pairing 
\bea 
(a,b) := \chi(X,a\otimes b)=\int_X \text{Td}(T_X)\text{ch}(a)\text{ch}(b)
\eea
it can be shown that $\mathcal{K}$ is a {\it Lagrangian} space with {\it Lagrangian polarization}  $\mathcal{K}=\mathcal{K}_+\oplus \mathcal{K}_-$.\\
The big J-function, $J(t)$, evaluated at $t=0$ ,$J(0)$, is called the {\it small} J-function.\\
For the computation we can re-write the 4d J-function in the following form
\bea 
J^{4d}&=&\frac{\prod_{j=1}^{5j_1}\prod_{k=1}^{5k_1}(1-\lambda_+\tilde{P}^{-j+5\epsilon}q^{-k})}{\prod_{j=1}^{j_1}\prod_{k=1}^{k_1}(1-\lambda_-\tilde{P}^{-j+\epsilon}q^{-k})^5}\frac{\prod_{j=1}^{5j_1}\prod_{k=0}^{\infty}(1-\lambda_+\tilde{P}^{-j+5\epsilon}q^{k})}{\prod_{j=1}^{j_1}\prod_{k=0}^{\infty}(1-\lambda_-\tilde{P}^{-j+\epsilon}q^{k})^5}\frac{\prod_{j=0}^{\infty}\prod_{k=1}^{5k_1}(1-\lambda_+\tilde{P}^{j+5\epsilon}q^{-k})}{\prod_{j=0}^{\infty}\prod_{k=1}^{k_1}(1-\lambda_-\tilde{P}^{j+\epsilon}q^{-k})^5}\nonumber\\&\times&
\frac{\prod_{j=1}^{j_1}\prod_{k=1}^{k_1}(1-\lambda_3\tilde{P}^{j+\epsilon}q^{k})^5}{\prod_{j=1}^{j_1}\prod_{k=1}^{\infty}(1-\lambda_3\tilde{P}^{j+\epsilon}q^{k})^5}\frac{\prod_{j=1}^{5j_1}\prod_{k=1}^{\infty}(1-\lambda_4\tilde{P}^{j+5\epsilon}q^{k})}{\prod_{j=1}^{5j_1}\prod_{k=1}^{5k_1}(1-\lambda_4\tilde{P}^{j+5\epsilon}q^{k})}\frac{\prod_{j=1}^{\infty}\prod_{k=1}^{5k_1}(1-\lambda_4\tilde{P}^{j+5\epsilon}q^{k})}{\prod_{j=1}^{\infty}\prod_{k=1}^{k_1}(1-\lambda_4\tilde{P}^{j+\epsilon}q^{k})^5}\nonumber\\&\times&
\frac{\prod_{j=0}^{\infty}\prod_{k=0}^{\infty}(1-\lambda_+\tilde{P}^{j+5\epsilon}q^{k})}{\prod_{j=0}^{\infty}\prod_{k=0}^{\infty}(1-\lambda_-\tilde{P}^{j+\epsilon}q^{k})^5}\frac{\prod_{j=1}^{\infty}\prod_{k=1}^{\infty}(1-\lambda_3\tilde{P}^{j+\epsilon}q^{k})^5}{\prod_{j=1}^{\infty}\prod_{k=1}^{\infty}(1-\lambda_4\tilde{P}^{j+5\epsilon}q^{k})}
\eea
The expression in the last line is taken as overall normalization factor\footnote{While performing computations on Mathematica we have worked with the normalized expression.}.

\section{Derivation of 4d difference operator}\label{Sec:Diff-Eqn}
In two dimensions, an elegant  way to describe the dependence of 2d supersymmetric partition function, also called vortex index, on $Q$ is in terms of a system of differential equations. For Calabi-Yau target spaces, these are the well-known Picard-Fuchs equations, whose solutions correspond to the periods of mirror manifold. The 3d counterpart of the Picard-Fuchs equation was derived in \cite{Jockers:2018sfl}. This 3d analogue involves certain finite difference operator and the solution set comprises the q-periods.

\subsection*{3d difference equation}
To elucidate the derivation of finite difference equations, the idea is to write down the integrand of the 3d index as a product of two factors. The recursion relation satisfied by the Q-dependent factor is the seed of the difference equation followed by the 3d index. Concretely, the 3d index can be given the following form \cite{Jockers:2018sfl}:
\bea 
Z=\text{ln}q\int \frac{d\epsilon}{2\pi i}f_{D^2}(q,\epsilon)I(Q,q,\epsilon),
\eea
where the Q-dependent part is given by
\bea 
I(Q,q,\epsilon)=\sum_{k\ge 0}Q^{k-\epsilon}a_k(q,\epsilon).
\eea
The $a_k(q,\epsilon)$ coefficients are given by
\bea 
a_k(q,\epsilon)=\frac{(-1)^{c_1 k}}{(q-1)^{c_1\epsilon}}\frac{\Gamma_q(1-\epsilon)^N\Gamma_q(1+l(k-\epsilon))}{\Gamma_q(1-l\epsilon)\Gamma_q(1+k-\epsilon)^N}
\eea
and satisfy the following recursion relation
\bea 
a_{k+1}=(-(1-q))^{c_1}\frac{\prod_{i=1}^l(1-q^{l(k-\epsilon)+i})}{(1-q^{k+1-\epsilon})^N}a_k.
\eea
One can verify this by following the properties of q-Gamma function $\Gamma_q(\epsilon)$.
By noting that 
\bea 
(1-q^{a Q\partial_Q+b})Q^{k-\epsilon}=(1-q^{a(k-\epsilon)+b})Q^{k-\epsilon},
\eea
one arrives at the final form of the 3d difference equation 
\bea 
\mathcal{L}I(Q,q,\epsilon)=0,\quad
\eea
where the operator $\mathcal{L}$ is defined by
\bea 
\mathcal{L}=(1-q^{Q\partial_Q})^N-(-1)^{c_1}Q\prod_{j=1}^l(1-q^{lQ\partial_Q+j}).
\eea

\subsection*{4d difference equation}
The 4d difference equation can be obtained by studying the coefficients in the $Q_1$, $Q_2$ expansion of the $J$-function given by
\bea
J^{4d}&=&\sum_{j_1=0}^{\infty}\sum_{k_1=0}^{\infty}Q_1^{j_1}Q_2^{k_1}e^{-i\frac{\pi}{3}(120a(-\sigma j_1+\epsilon\sigma-\tau k_1)^3+20b(-\sigma j_1+\epsilon\sigma-\tau k_1)^2)}\nonumber\\&\times&
\prod_{j=0}^{\infty}\prod_{k=0}^{\infty}\frac{1-\tilde{P}^{j-5j_1+5\epsilon}q^{k-5k_1}}{(1-\tilde{P}^{j-j_1+\epsilon}q^{k-k_1})^5}\prod_{j=0}^{\infty}\prod_{k=0}^{\infty}\frac{(1-\tilde{P}^{j+j_1+1-\epsilon}q^{k+k_1+1})^5}{1-\tilde{P}^{j+5j_1+1-5\epsilon}q^{k+5k_1+1}}.
\eea
Following the logic of the previous subsection, we arrive at the following 4d difference operator 

\bea\label{eq:Diffeq}
&&\mathcal{L}^{4d}=\nonumber\\&&\Big\{\prod_{j=-1}^{\infty}(1-p_pp_q\tilde{P}^{j+1-\epsilon})^5\prod_{k=0}^{\infty}(1-p_pp_q\tilde{P}^{-\epsilon}q^{k+1})^5 \prod_{k=1}^{\infty}(1-p_p^{-1}p_q^{-1}\tilde{P}^{\epsilon}q^{k})^5\prod_{j=0}^{\infty}(1-p_p^{-1}p_q^{-1}\tilde{P}^{j+\epsilon})^5\Big\}\nonumber\\&-&
\Big\{e^{-i\frac{\pi}{3}\Big(-120 a(\sigma+\tau)^3-360 a(\sigma+\tau)(-\sigma ln(p_p)+\epsilon\sigma-\tau ln(p_q))^2+(360a(\sigma+\tau)^2-40b(\sigma+\tau))(-\sigma ln(p_p)+\epsilon\sigma-\tau ln(p_q))+20b(\sigma+\tau)^2  \Big)}\nonumber\\
&\times&Q_1Q_2\prod_{j=1}^5\prod_{k=1}^5(1-p_p^{-5}p_q^{-5}\tilde{P}^{-j+5\epsilon}q^{-k})\prod_{j=1}^5\prod_{k=0}^{\infty}(1-p_p^{-5}p_q^{-5}\tilde{P}^{-j+5\epsilon}q^{k})\nonumber\\&\times&\prod_{j=0}^{\infty}\prod_{k=1}^5(1-p_p^{-5}p_q^{-5}\tilde{P}^{j+5\epsilon}q^{-k})\prod_{j=0}^{\infty}\prod_{k=0}^4(1-p_p^{5}p_q^{5}\tilde{P}^{j-1-5\epsilon}q^{k+1})\prod_{j=0}^4\prod_{k=5}^{\infty}(1-p_p^{5}p_q^{5}\tilde{P}^{j+1-5\epsilon}q^{k+1})\Big\},\nonumber\\
\eea
where $p_p=\tilde{P}^{Q_1\partial_{Q_1}}$ and $p_q=q^{Q_2\partial_{Q_2}}$.
In gauge theory interpretation, the fugacities $Q_1$ and $Q_2$ originate from evaluating the classical 4d $\mathcal{N}=1$ action on the localization locus.\\ Next we provide a proof that $\mathcal{L}^{4d}$ indeed annihilates $J^{4d}$.
\subsection*{Proof}
Acting with the difference operator of Eq. (\ref{eq:Diffeq}) on the J-function, $J^{4d}$, we get
\bea\label{eq:Diffeq2}
&&\mathcal{L}^{4d}J^{4d}=\nonumber\\
&&\Big\{\prod_{j=-1}^{\infty}(1-p_pp_q\tilde{P}^{j+1-\epsilon})^5\prod_{k=0}^{\infty}(1-p_pp_q\tilde{P}^{-\epsilon}q^{k+1})^5 \prod_{k=1}^{\infty}(1-p_p^{-1}p_q^{-1}\tilde{P}^{\epsilon}q^{k})^5\nonumber\\&&\times \prod_{j=0}^{\infty}(1-p_p^{-1}p_q^{-1}\tilde{P}^{j+\epsilon})^5 \Big(\sum_{j_1=0}^{\infty}\sum_{k_1=0}^{\infty}Q_1^{j_1}Q_2^{k_1}e^{-i\frac{\pi}{3}(120a(-\sigma j_1+\epsilon\sigma-\tau k_1)^3+20b(-\sigma j_1+\epsilon\sigma-\tau k_1)^2)}\nonumber\\&&\times
\prod_{j=0}^{\infty}\prod_{k=0}^{\infty}\frac{1-\tilde{P}^{j-5j_1+5\epsilon}q^{k-5k_1}}{(1-\tilde{P}^{j-j_1+\epsilon}q^{k-k_1})^5}\prod_{j=0}^{\infty}\prod_{k=0}^{\infty}\frac{(1-\tilde{P}^{j+j_1+1-\epsilon}q^{k+k_1+1})^5}{1-\tilde{P}^{j+5j_1+1-5\epsilon}q^{k+5k_1+1}}\Big)\Big\}\nonumber\\&-&
\Big\{e^{-i\frac{\pi}{3}\Big(-120 a(\sigma+\tau)^3-360 a(\sigma+\tau)(-\sigma ln(p_p)+\epsilon\sigma-\tau ln(p_q))^2+(360a(\sigma+\tau)^2-40b(\sigma+\tau))(-\sigma ln(p_p)+\epsilon\sigma-\tau ln(p_q))+20b(\sigma+\tau)^2  \Big)}\nonumber\\
&\times&Q_1Q_2\prod_{j=1}^5\prod_{k=1}^5(1-p_p^{-5}p_q^{-5}\tilde{P}^{-j+5\epsilon}q^{-k})\prod_{j=1}^5\prod_{k=0}^{\infty}(1-p_p^{-5}p_q^{-5}\tilde{P}^{-j+5\epsilon}q^{k})\nonumber\\&\times&\prod_{j=0}^{\infty}\prod_{k=1}^5(1-p_p^{-5}p_q^{-5}\tilde{P}^{j+5\epsilon}q^{-k})\prod_{j=0}^{\infty}\prod_{k=0}^4(1-p_p^{5}p_q^{5}\tilde{P}^{j-1-5\epsilon}q^{k+1})\prod_{j=0}^4\prod_{k=5}^{\infty}(1-p_p^{5}p_q^{5}\tilde{P}^{j+1-5\epsilon}q^{k+1})\nonumber\\&&\times \Big(\sum_{j_1=0}^{\infty}\sum_{k_1=0}^{\infty}Q_1^{j_1}Q_2^{k_1}e^{-i\frac{\pi}{3}(120a(-\sigma j_1+\epsilon\sigma-\tau k_1)^3+20b(-\sigma j_1+\epsilon\sigma-\tau k_1)^2)}\nonumber\\&&\times
\prod_{j=0}^{\infty}\prod_{k=0}^{\infty}\frac{1-\tilde{P}^{j-5j_1+5\epsilon}q^{k-5k_1}}{(1-\tilde{P}^{j-j_1+\epsilon}q^{k-k_1})^5}\prod_{j=0}^{\infty}\prod_{k=0}^{\infty}\frac{(1-\tilde{P}^{j+j_1+1-\epsilon}q^{k+k_1+1})^5}{1-\tilde{P}^{j+5j_1+1-5\epsilon}q^{k+5k_1+1}}\Big)\Big\}\nonumber\\
\eea
\bea
&&\mathcal{L}^{4d}J^{4d}= \Big\{\sum_{j_1=0}^{\infty}\sum_{k_1=0}^{\infty}Q_1^{j_1}Q_2^{k_1}e^{-i\frac{\pi}{3}(120a(-\sigma j_1+\epsilon\sigma-\tau k_1)^3+20b(-\sigma j_1+\epsilon\sigma-\tau k_1)^2)}\nonumber\\
&&\prod_{j=-1}^{\infty}(1-\tilde{P}^{j+j_1+1-\epsilon}q^{k_1})^5\prod_{k=0}^{\infty}(1-\tilde{P}^{j_1-\epsilon}q^{k+k_1+1})^5 \prod_{k=1}^{\infty}(1-\tilde{P}^{-j_1+\epsilon}q^{-k_1+k})^5\nonumber\\&&\times\prod_{j=0}^{\infty}(1-\tilde{P}^{-j_1+j+\epsilon}q^{-k_1})^5
\prod_{j=0}^{\infty}\prod_{k=0}^{\infty}\frac{1-\tilde{P}^{j-5j_1+5\epsilon}q^{k-5k_1}}{(1-\tilde{P}^{j-j_1+\epsilon}q^{k-k_1})^5}\prod_{j=0}^{\infty}\prod_{k=0}^{\infty}\frac{(1-\tilde{P}^{j+j_1+1-\epsilon}q^{k+k_1+1})^5}{1-\tilde{P}^{j+5j_1+1-5\epsilon}q^{k+5k_1+1}}\Big\}\nonumber\\&-&
 \Big\{\sum_{j_1=0}^{\infty}\sum_{k_1=0}^{\infty}Q_1^{j_1+1}Q_2^{k_1+1}e^{-i\frac{\pi}{3}(120a(-\sigma (j_1+1)+\epsilon\sigma-\tau (k_1+1))^3+20b(-\sigma (j_1+1)+\epsilon\sigma-\tau (k_1+1))^2)}\nonumber\\
&\times&\prod_{j=1}^5\prod_{k=1}^5(1-\tilde{P}^{-5j_1-j+5\epsilon}q^{-5k_1-k})\prod_{j=1}^5\prod_{k=0}^{\infty}(1-\tilde{P}^{-5j_1-j+5\epsilon}q^{-5k_1+k})\nonumber\\&\times&\prod_{j=0}^{\infty}\prod_{k=1}^5(1-\tilde{P}^{-5j_1+j+5\epsilon}q^{-5k_1-k})\prod_{j=0}^{\infty}\prod_{k=0}^4(1-\tilde{P}^{5j_1+j-1-5\epsilon}q^{5k_1+k+1})\nonumber\\&&\times\prod_{j=0}^4\prod_{k=5}^{\infty}(1-\tilde{P}^{5j_1+j+1-5\epsilon}q^{5k_1+k+1})\nonumber\\&&\times
\prod_{j=0}^{\infty}\prod_{k=0}^{\infty}\frac{1-\tilde{P}^{j-5j_1+5\epsilon}q^{k-5k_1}}{(1-\tilde{P}^{j-j_1+\epsilon}q^{k-k_1})^5}\prod_{j=0}^{\infty}\prod_{k=0}^{\infty}\frac{(1-\tilde{P}^{j+j_1+1-\epsilon}q^{k+k_1+1})^5}{1-\tilde{P}^{j+5j_1+1-5\epsilon}q^{k+5k_1+1}}\Big\}.\nonumber\\
\eea
In the second term of the last expression let us change the dummy indices $j_1,k_1$ to $j_1^{\prime}=j_1-1$ and $k_1^{\prime}=k_1-1$, respectively. As a result, we obtain

\bea\label{eq:diffeq}
&&\mathcal{L}^{4d}J^{4d}=
\Big\{\sum_{j_1=0}^{\infty}\sum_{k_1=0}^{\infty}Q_1^{j_1}Q_2^{k_1}e^{-i\frac{\pi}{3}(120a(-\sigma j_1+\epsilon\sigma-\tau k_1)^3+20b(-\sigma j_1+\epsilon\sigma-\tau k_1)^2)}\nonumber\\
&&\prod_{j=-1}^{\infty}(1-\tilde{P}^{j+j_1+1-\epsilon}q^{k_1})^5\prod_{k=0}^{\infty}(1-\tilde{P}^{j_1-\epsilon}q^{k+k_1+1})^5 \prod_{k=1}^{\infty}(1-\tilde{P}^{-j_1+\epsilon}q^{-k_1+k})^5\nonumber\\&&\times\prod_{j=0}^{\infty}(1-\tilde{P}^{-j_1+j+\epsilon}q^{-k_1})^5
\prod_{j=0}^{\infty}\prod_{k=0}^{\infty}\frac{1-\tilde{P}^{j-5j_1+5\epsilon}q^{k-5k_1}}{(1-\tilde{P}^{j-j_1+\epsilon}q^{k-k_1})^5}\prod_{j=0}^{\infty}\prod_{k=0}^{\infty}\frac{(1-\tilde{P}^{j+j_1+1-\epsilon}q^{k+k_1+1})^5}{1-\tilde{P}^{j+5j_1+1-5\epsilon}q^{k+5k_1+1}}\Big\}\nonumber\\&-&
\Big\{ \sum_{j_1=1}^{\infty}\sum_{k_1=1}^{\infty}Q_1^{j_1}Q_2^{k_1}e^{-i\frac{\pi}{3}(120a(-\sigma (j_1)+\epsilon\sigma-\tau (k_1))^3+20b(-\sigma (j_1)+\epsilon\sigma-\tau (k_1))^2)}\nonumber\\
&\times&\prod_{j=1}^5\prod_{k=1}^5(1-\tilde{P}^{-5j_1+5-j+5\epsilon}q^{-5k_1+5-k})\prod_{j=1}^5\prod_{k=0}^{\infty}(1-\tilde{P}^{-5j_1+5-j+5\epsilon}q^{-5k_1+5+k})\nonumber\\&\times&\prod_{j=0}^{\infty}\prod_{k=1}^5(1-\tilde{P}^{-5j_1+5+j+5\epsilon}q^{-5k_1+5-k})\prod_{j=0}^{\infty}\prod_{k=0}^4(1-\tilde{P}^{5j_1-5+j-1-5\epsilon}q^{5k_1-5+k+1})\nonumber\\&&\times\prod_{j=0}^4\prod_{k=5}^{\infty}(1-\tilde{P}^{5j_1-5+j+1-5\epsilon}q^{5k_1-5+k+1})\nonumber\\&&\times
\prod_{j=0}^{\infty}\prod_{k=0}^{\infty}\frac{1-\tilde{P}^{j-5j_1+5+5\epsilon}q^{k-5k_1+5}}{(1-\tilde{P}^{j-j_1+1+\epsilon}q^{k-k_1+1})^5}\prod_{j=0}^{\infty}\prod_{k=0}^{\infty}\frac{(1-\tilde{P}^{j+j_1-1+1-\epsilon}q^{k+k_1-1+1})^5}{1-\tilde{P}^{j+5j_1-5+1-5\epsilon}q^{k+5k_1-5+1}}\Big\}.\nonumber\\
\eea

Observe the following identities 
\bea 
&&\prod_{j=-5}^{\infty}\prod_{k=-5}^{\infty}(1-\tilde{P}^{j-5j_1+5+5\epsilon}q^{k-5k_1+5})=\nonumber\\
&&\prod_{j=1}^{5}\prod_{k=1}^{5}(1-\tilde{P}^{-j-5j_1+5+5\epsilon}q^{-k-5k_1+5})\prod_{j=0}^{\infty}\prod_{k=1}^{5}(1-\tilde{P}^{j-5j_1+5+5\epsilon}q^{-k-5k_1+5})\nonumber\\
&\times& \prod_{j=1}^{5}\prod_{k=0}^{\infty}(1-\tilde{P}^{-j-5j_1+5+5\epsilon}q^{k-5k_1+5})\prod_{j=0}^{\infty}\prod_{k=0}^{\infty}(1-\tilde{P}^{j-5j_1+5+5\epsilon}q^{k-5k_1+5})\nonumber\\
&&\prod_{j=-5}^{\infty}\prod_{k=-5}^{\infty}(1-\tilde{P}^{j-5j_1+5+5\epsilon}q^{k-5k_1+5})=\prod_{j=0}^{\infty}\prod_{k=0}^{\infty}(1-\tilde{P}^{j-5j_1+5\epsilon}q^{k-5k_1})\nonumber\\
\eea
\bea 
&&\prod_{j=0}^{\infty}\prod_{k=0}^{\infty}(1-\tilde{P}^{j+5j_1-5+1-5\epsilon}q^{k+5k_1-5+1})=\nonumber\\&&\Big(\prod_{j=0}^{\infty}\prod_{k=0}^{4}(1-\tilde{P}^{j+5j_1-5+1-5\epsilon}q^{k+5k_1-5+1})\prod_{j=0}^{4}\prod_{k=5}^{\infty}(1-\tilde{P}^{j+5j_1-5+1-5\epsilon}q^{k+5k_1-5+1})\nonumber\\&\times&\prod_{j=5}^{\infty}\prod_{k=5}^{\infty}(1-\tilde{P}^{j+5j_1-5+1-5\epsilon}q^{k+5k_1-5+1})\prod_{j=5}^{\infty}\prod_{k=5}^{\infty}(1-\tilde{P}^{j+5j_1-5+1-5\epsilon}q^{k+5k_1-5+1})\Big)\nonumber\\
&&=\prod_{j=0}^{\infty}\prod_{k=0}^{\infty}(1-\tilde{P}^{j+5j_1+1-5\epsilon}q^{k+5k_1+1})\nonumber\\
\eea
\bea 
&&\prod_{j=0}^{\infty}\prod_{k=0}^{\infty}(1-\tilde{P}^{j+j_1-1+1-\epsilon}q^{k+k_1-1+1})=\prod_{j=-1}^{\infty}\prod_{k=-1}^{\infty}(1-\tilde{P}^{j+j_1+1+1-\epsilon}q^{k+k_1+1+1})\nonumber\\&&=\prod_{j=-1}^{\infty}(1-\tilde{P}^{j+j_1+1-\epsilon}q^{k_1})\prod_{k=0}^{\infty}(1-\tilde{P}^{j_1-\epsilon}q^{k+k_1+1})\prod_{j=0}^{\infty}\prod_{k=0}^{\infty}(1-\tilde{P}^{j+j_1+1-\epsilon}q^{k+k_1+1})\nonumber\\
\eea
\bea
&&\prod_{j=0}^{\infty}\prod_{k=0}^{\infty}(1-\tilde{P}^{j-j_1+1+\epsilon}q^{k-k_1+1})=\frac{\prod_{j=0}^{\infty}\prod_{k=0}^{\infty}(1-\tilde{P}^{j-j_1+\epsilon}q^{k-k_1})}{\prod_{k=1}^{\infty}(1-\tilde{P}^{-j_1+\epsilon}q^{k-k_1})\prod_{j=0}^{\infty}(1-\tilde{P}^{j-j_1+\epsilon}q^{-k_1})}.\nonumber\\
\eea
Using these identities in Eq. (\ref{eq:diffeq}) we get

\bea
&&\mathcal{L}^{4d}J^{4d}=\Big\{ \sum_{j_1=0}^{\infty}\sum_{k_1=0}^{\infty}Q_1^{j_1}Q_2^{k_1}e^{-i\frac{\pi}{3}(120a(-\sigma j_1+\epsilon\sigma-\tau k_1)^3+20b(-\sigma j_1+\epsilon\sigma-\tau k_1)^2)}\nonumber\\
&&\prod_{j=-1}^{\infty}(1-\tilde{P}^{j+j_1+1-\epsilon}q^{k_1})^5\prod_{k=0}^{\infty}(1-\tilde{P}^{j_1-\epsilon}q^{k+k_1+1})^5 \prod_{k=1}^{\infty}(1-\tilde{P}^{-j_1+\epsilon}q^{-k_1+k})^5\nonumber\\&&\times\prod_{j=0}^{\infty}(1-\tilde{P}^{-j_1+j+\epsilon}q^{-k_1})^5
\prod_{j=0}^{\infty}\prod_{k=0}^{\infty}\frac{1-\tilde{P}^{j-5j_1+5\epsilon}q^{k-5k_1}}{(1-\tilde{P}^{j-j_1+\epsilon}q^{k-k_1})^5}\prod_{j=0}^{\infty}\prod_{k=0}^{\infty}\frac{(1-\tilde{P}^{j+j_1+1-\epsilon}q^{k+k_1+1})^5}{1-\tilde{P}^{j+5j_1+1-5\epsilon}q^{k+5k_1+1}}\Big\}\nonumber\\&-&
\Big\{ \sum_{j_1=1}^{\infty}\sum_{k_1=1}^{\infty}Q_1^{j_1}Q_2^{k_1}e^{-i\frac{\pi}{3}(120a(-\sigma j_1+\epsilon\sigma-\tau k_1)^3+20b(-\sigma j_1+\epsilon\sigma-\tau k_1)^2)}\nonumber\\
&&\prod_{j=-1}^{\infty}(1-\tilde{P}^{j+j_1+1-\epsilon}q^{k_1})^5\prod_{k=0}^{\infty}(1-\tilde{P}^{j_1-\epsilon}q^{k+k_1+1})^5 \prod_{k=1}^{\infty}(1-\tilde{P}^{-j_1+\epsilon}q^{-k_1+k})^5\nonumber\\
&&\times\prod_{j=0}^{\infty}(1-\tilde{P}^{-j_1+j+\epsilon}q^{-k_1})^5
\prod_{j=0}^{\infty}\prod_{k=0}^{\infty}\frac{1-\tilde{P}^{j-5j_1+5\epsilon}q^{k-5k_1}}{(1-\tilde{P}^{j-j_1+\epsilon}q^{k-k_1})^5}\prod_{j=0}^{\infty}\prod_{k=0}^{\infty}\frac{(1-\tilde{P}^{j+j_1+1-\epsilon}q^{k+k_1+1})^5}{1-\tilde{P}^{j+5j_1+1-5\epsilon}q^{k+5k_1+1}}\Big\}.\nonumber\\
\eea
Notice that in the last expression the first bracketed term exactly cancels the second bracketed term except for the $j_1=0,k_1=0$ term.
Simplifying the last expression, we are left with

\bea
&&\mathcal{L}^{4d}J^{4d}= e^{-i\frac{\pi}{3}(120a(\epsilon\sigma+\eta\tau)^3+20b(+\epsilon\sigma)^2)}\nonumber\\
&&\prod_{j=-1}^{\infty}(1-\tilde{P}^{j+1-\epsilon})^5\prod_{k=0}^{\infty}(1-\tilde{P}^{-\epsilon}q^{k+1})^5 \prod_{k=1}^{\infty}(1-\tilde{P}^{+\epsilon}q^{k})^5\prod_{j=0}^{\infty}(1-\tilde{P}^{+j+\epsilon})^5\nonumber\\&&\times
\prod_{j=0}^{\infty}\prod_{k=0}^{\infty}\frac{1-\tilde{P}^{j+5\epsilon}q^{k}}{(1-\tilde{P}^{j+\epsilon}q^{k})^5}\prod_{j=0}^{\infty}\prod_{k=0}^{\infty}\frac{(1-\tilde{P}^{j+1-\epsilon}q^{k+1})^5}{1-\tilde{P}^{j+1-5\epsilon}q^{k+1}}\nonumber\\&&=
\frac{(1-\tilde{P}^{-\epsilon})^5(1-\tilde{P}^{\epsilon})^5}{(1-\tilde{P}^{\epsilon})^5} e^{-i\frac{\pi}{3}(120a(\epsilon\sigma)^3+20b(+\epsilon\sigma)^2)}\nonumber\\
&\times&\prod_{j=0}^{\infty}(1-\tilde{P}^{j+1-\epsilon})^5\prod_{k=0}^{\infty}(1-\tilde{P}^{-\epsilon}q^{k+1})^5 \prod_{k=1}^{\infty}(1-\tilde{P}^{+\epsilon}q^{k})^5\prod_{j=1}^{\infty}(1-\tilde{P}^{+j+\epsilon})^5\nonumber\\&&\times
\prod_{j=0}^{\infty}\prod_{k=0,j=k\neq 0}^{\infty}\frac{1-\tilde{P}^{j+5\epsilon}q^{k}}{(1-\tilde{P}^{j+\epsilon}q^{k})^5}\prod_{j=0}^{\infty}\prod_{k=0}^{\infty}\frac{(1-\tilde{P}^{j+1-\epsilon}q^{k+1})^5}{1-\tilde{P}^{j+1-5\epsilon}q^{k+1}}.\nonumber\\
\eea
In K-theory, however, we have 
\bea \label{eq:PH}
(1-\tilde{P}^{-\epsilon})^5=(1-P)^5=0,
\eea
where $P=\tilde{P}^{-\epsilon}$. Therefore, 
\bea 
\mathcal{L}^{4d}J^{4d}=0.
\eea
\section{First degeneration limit: $q \to 0$}\label{Sec:qto0}
The two limits $q \to 0$ and $q\to 1$ correspond to $\tau\to \infty$ and $\tau \to 0$ respectively. This reminds one of the $SL(2,\mathbb{Z})$ symmetry under the transformation $\tau\to-\frac{1}{\tau}$. 
A natural degeneration limit to consider is $q\to 0$ limit. In this limit the elliptic gamma function reduces to the 3d gamma function. The elliptic gamma function is defined by
\bea 
\Gamma_e(x;p,q)=\frac{\prod_{j,k\ge 0 }(1-p^{j+1}q^{k+1}x^{-1})}{\prod_{j,k\ge 0 }(1-p^{j}q^{k}x)}.
\eea
Note that as $q\to 0$
\bea 
\lim_{q\to 0}\Gamma_e(x;p,q)=\Gamma(u,\sigma)=\prod_{n\ge 0}(1-e^{2\pi i(u+n\sigma)})^{-1},
\eea
where $x=e^{2\pi i u}$ and $p=e^{2\pi i \sigma}$. The gamma function $\Gamma(u,\sigma)$ appears in one loop determinants of a 3d gauge theory on $D^2\times \mathbb{S}^1$.
\subsection{$q \to 0$ limit of the 4d partition function}
It would be a useful exercise to study the $q \to 0$ limit of the 4d J-function and the 4d difference equation.
The 4d J-function coming from the normalized partition function (\ref{eq:quinticnorma}) has the following form
\bea
J^{4d}&=&\sum_{j_1\ge 0}Q_1^{j_1}e^{\frac{- i \pi }{3}(\frac{120}{\tau\sigma}(-\sigma j_1+\epsilon \sigma)^3-\frac{30}{\tau\sigma}(1+\tau+\sigma)(-\sigma j_1+\epsilon \sigma)^2)}\nonumber\\
&\times&\frac{\prod_{j=1}^{5j_1}\prod_{k=0}^{\infty}(1-\lambda_+\tilde{P}^{-j+5\epsilon}q^k)\prod_{j=1}^{5j_1}\prod_{k=1}^{\infty}(1-\lambda_4\tilde{P}^{j+5\epsilon}q^k)}{\prod_{j=1}^{j_1}\prod_{k=0}^{\infty}(1-\lambda_-\tilde{P}^{-j+\epsilon}q^k)^5\prod_{j=1}^{j_1}\prod_{k=1}^{\infty}(1-\lambda_3\tilde{P}^{j+\epsilon}q^k)^5}\nonumber\\
&+&\sum_{j_1\ge 0}\sum_{k_1\ge 1}Q_1^{j_1}Q_2^{k_1}e^{\frac{- i \pi }{3}(\frac{120}{\tau\sigma}(-\sigma j_1+\epsilon \sigma-\tau k_1)^3-\frac{30}{\tau\sigma}(1+\tau+\sigma)(-\sigma j_1+\epsilon \sigma-\tau k_1)^2)}\nonumber\\
&\times& \prod_{j=0}^{\infty}\prod_{k=0}^{\infty}\frac{(1-\lambda_+\tilde{P}^{j-5j_1+5\epsilon}q^{k-5k_1})}{(1-\lambda_-\tilde{P}^{j-j_1+\epsilon}q^{k-k_1})^5}\prod_{j=0}^{\infty}\prod_{k=0}^{\infty}\frac{(1-\lambda_4\tilde{P}^{j+j_1+\epsilon+1}q^{k+k_1+1})}{(1-\lambda_3\tilde{P}^{j+5j_1+5\epsilon+1}q^{k+5k_1+1})^5}.
\eea
Let us consider the limit $q\to 0$ or $\tau\to \pm i\infty$.  In this limit, the 
partition function simplifies to
\bea
J^{4d}&=&\sum_{j_1\ge 0}Q_1^{j_1}e^{\frac{- i \pi }{3}(-\frac{30}{\sigma}(-\sigma j_1+\epsilon \sigma)^2)}
\frac{\prod_{j=1}^{5j_1}(1-\lambda_+\tilde{P}^{-j+5\epsilon})}{\prod_{j=1}^{j_1}(1-\lambda_-\tilde{P}^{-j+\epsilon})^5}\nonumber\\
&+&\sum_{j_1\ge 0}\sum_{k_1\ge 1}Q_1^{j_1}Q_2^{k_1}e^{\frac{- i \pi }{3}(-\frac{120}{\sigma}\tau^2 k_1^3-\frac{30}{\sigma}(\tau k_1)^2)}\nonumber\\
&\times& \lim_{q\to 0} \prod_{j=0}^{\infty}\prod_{k=0}^{\infty}\frac{(1-\lambda_+\tilde{P}^{j-5j_1+5\epsilon}q^{k-5k_1})}{(1-\lambda_-\tilde{P}^{j-j_1+\epsilon}q^{k-k_1})^5} \lim_{q\to 0}\prod_{j=0}^{\infty}\prod_{k=0}^{\infty}\frac{(1-\lambda_4\tilde{P}^{j+j_1+\epsilon+1}q^{k+k_1+1})}{(1-\lambda_3\tilde{P}^{j+5j_1+5\epsilon+1}q^{k+5k_1+1})^5}.\nonumber\\
\eea
The second limit in the last line evaluates to $1$. The first limit requires some care.
\bea
&&\lim_{q\to 0} \prod_{j=0}^{\infty}\prod_{k=0}^{\infty}\frac{(1-\lambda_+\tilde{P}^{j-5j_1+5\epsilon}q^{k-5k_1})}{(1-\lambda_-\tilde{P}^{j-j_1+\epsilon}q^{k-k_1})^5}=\lim_{q\to 0} \frac{\prod_{j=0}^{\infty}\prod_{k=0}^{5k_1-1}(1-\lambda_+\tilde{P}^{j-5j_1+5\epsilon}q^{k-5k_1})}{\prod_{j=0}^{\infty}\prod_{k=0}^{k_1-1}(1-\lambda_-\tilde{P}^{j-j_1+\epsilon}q^{k-k_1})^5} \nonumber\\
&\times&  \frac{\prod_{j=0}^{\infty}(1-\lambda_+\tilde{P}^{j-5j_1+5\epsilon})}{\prod_{j=0}^{\infty}(1-\lambda_-\tilde{P}^{j-j_1+\epsilon})^5}\lim_{q\to 0} \frac{\prod_{j=0}^{\infty}\prod_{k=5k_1+1}^{\infty}(1-\lambda_+\tilde{P}^{j-5j_1+5\epsilon}q^{k-5k_1})}{\prod_{j=0}^{\infty}\prod_{k=k_1+1}^{\infty}(1-\lambda_-\tilde{P}^{j-j_1+\epsilon}q^{k-k_1})^5}.
\eea
The third factor on the right hand side in the last equation evaluates to $1$. To analyze the first factor, we can write it as follows
\bea 
&&\lim_{q\to 0} \frac{\prod_{j=0}^{\infty}\prod_{k=0}^{5k_1-1}(1-\lambda_+\tilde{P}^{j-5j_1+5\epsilon}q^{k-5k_1})}{\prod_{j=0}^{\infty}\prod_{k=0}^{k_1-1}(1-\lambda_-\tilde{P}^{j-j_1+\epsilon}q^{k-k_1})^5}=\lim_{q\to 0} \frac{\prod_{j=0}^{\infty}\prod_{k=0}^{5k_1-1}(\frac{1}{q^{5k_1-k}})(q^{5k_1-k}-\lambda_+\tilde{P}^{j-5j_1+5\epsilon}q^{k-5k_1})}{\prod_{j=0}^{\infty}\prod_{k=0}^{k_1-1}(\frac{1}{q^{5(k_1-k)}})(q^{(k_1-k)}-\lambda_-\tilde{P}^{j-j_1+\epsilon}q^{k-k_1})^5}\nonumber\\
&=&\lim_{q\to 0} \frac{\prod_{j=0}^{\infty}\prod_{k=0}^{5k_1-1}(\frac{1}{q^{5k_1-k}})}{\prod_{j=0}^{\infty}\prod_{k=0}^{k_1-1}(\frac{1}{q^{5(k_1-)k}})}
\lim_{q\to 0} \frac{\prod_{j=0}^{\infty}\prod_{k=0}^{5k_1-1}(q^{5k_1-k}-\lambda_+\tilde{P}^{j-5j_1+5\epsilon}q^{k-5k_1})}{\prod_{j=0}^{\infty}\prod_{k=0}^{k_1-1}(q^{(k_1-k)}-\lambda_-\tilde{P}^{j-j_1+\epsilon}q^{k-k_1})^5}.
\eea
The second factor again gives some non-zero finite number. The first factor can be simplified to
\bea 
\lim_{q\to 0} \frac{\prod_{j=0}^{\infty}\prod_{k=0}^{5k_1-1}(\frac{1}{q^{5k_1-k}})}{\prod_{j=0}^{\infty}\prod_{k=0}^{k_1-1}(\frac{1}{q^{5(k_1-)k}})}=\lim_{q\to 0}\prod_{j=0}^{\infty}q^{-10k_1^2}=\lim_{q\to 0}(q^{-10k_1^2})^{\sum_{j=0}^{\infty}1},
\eea
using Zeta-function regularization, we can substitute $\sum_{j=0}^{\infty}1=\zeta(0)=\frac{1}{2}$
in the last equation to get 
\bea
\lim_{q\to 0}(q^{-10k_1^2})^{\sum_{j=0}^{\infty}1}=\lim_{q\to 0}(q^{-5k_1^2}).
\eea
This leads to the final expression
\bea
J^{4d}&=&\sum_{j_1\ge 0}Q_1^{j_1}e^{\frac{- i \pi }{3}(-\frac{30}{\sigma}(-\sigma j_1+\epsilon \sigma)^2)}
\frac{\prod_{j=1}^{5j_1}(1-\lambda_+\tilde{P}^{-j+5\epsilon})}{\prod_{j=1}^{j_1}(1-\lambda_-\tilde{P}^{-j+\epsilon})^5}\nonumber\\
&+&\sum_{j_1\ge 0}\sum_{k_1\ge 1}Q_1^{j_1}Q_2^{k_1}\lim_{\tau\to i\infty}\Big(q^{-5k_1^2}e^{\frac{- i \pi }{3}(-\frac{120}{\sigma}\tau^2 k_1^3-\frac{30}{\sigma}(\tau k_1)^2)}\Big)\nonumber\\
&\times&  \prod_{j=0}^{\infty}\frac{(1-\lambda_+\tilde{P}^{j-5j_1+5\epsilon})}{(1-\lambda_-\tilde{P}^{j-j_1+\epsilon})^5}.
\eea
This should be compare to the 3d expression of the J-function \cite{Jockers:2018sfl}
\bea 
J^{3d}&=&\sum_{j_1\ge 0}
\frac{\prod_{j=1}^{5j_1}(1-\lambda_+\tilde{P}^{-j+5\epsilon})}{\prod_{j=1}^{j_1}(1-\lambda_-\tilde{P}^{-j+\epsilon})^5}.\nonumber\\ 
\eea
\subsection{$q \to 0$ limit of the 4d difference equation }
It is interesting to see how the 4d difference equation related to $D^2\times \mathbb{T}^2$ degenerates to 3d difference equation related to $D^2\times \mathbb{S}^1$. It turns out that in $q\to 0$ limit,the 4d difference equation reduces to a certain deformation of the 3d difference equation.
The $q\to 0$ limit  of the 4d difference operator acting on J-function Eq. (\ref{eq:Diffeq2}) turns out to be given by the following expression
\bea
&&\lim_{q\to 0}\mathcal{L}^{4d}J^{4d}=\nonumber\\
&&\Big\{\sum_{j_1=0}Q_1^{j_1}e^{-i\frac{\pi}{3}(-120a(\sigma \epsilon-\sigma j_1)^3+30b(-\sigma j_1+\sigma \epsilon)^2)}\prod_{j=0}^{\infty}(1-\tilde{p}^{j-j_1+\epsilon})^5(1-\tilde{p}^{j_1-\epsilon})^5
\prod_{j=0}^{\infty}(1-\tilde{p}^{j+j_1+1-\epsilon})^5\nonumber\\
&\times&\prod_{j=0}^{\infty}\frac{1-\tilde{p}^{j-5j_1+5\epsilon}}{(1-\tilde{p}^{j-j_1+\epsilon})^5}+\sum_{j_1=0}^{\infty}\sum_{k_1=1}^{\infty}e^{-i\frac{\pi}{3}(120a(\tau k_1)^3+30b(\tau k_1)^2)}(-1)^{5(k_1-1)}\tilde{p}^{5(k_1-1)(-j_1+\epsilon)-\frac{5}{12}-\frac{5 j_1}{2}+\frac{5\epsilon}{2}-10k_1\epsilon}q^{-5k_1^2}\nonumber\\&\times&(1-\tilde{p}^{-j_1+\epsilon})^5\prod_{j=0}^{\infty}\frac{1-\tilde{p}^{j-5j_1+5\epsilon}}{(1-\tilde{p}^{j-j_1+\epsilon})^5}Q_1^{j_1}Q_2^{k_1}\Big\}-\nonumber\\&&\Big\{\sum_{j_1=0}^{\infty}Q_1^{j_1+1}Q_2e^{-i\frac{\pi}{3}(-120a(\sigma \epsilon-\sigma (j_1+1))^3+30b(-\sigma (j_1+1)+\sigma \epsilon)^2)}(-1)^{\frac{5.11}{2}}\tilde{p}^{-\frac{31.5}{2}-\frac{25.11.j_1}{2}+\frac{25.11.\epsilon}{2}}q^{-\frac{15.11}{2}}\nonumber\\&\times&\prod_{j=0}^{\infty}\frac{1-\tilde{p}^{j-5j_1+5\epsilon}}{(1-\tilde{p}^{j-j_1+\epsilon})^5}+
\sum_{j_1=0}^{\infty}\sum_{k_1=1}^{\infty}Q_1^{j_1+1}Q_2^{k_1+1}e^{-i\frac{\pi}{3}(120a(\tau (k_1+1))^3+30b(\tau (k_1+1))^2)}(-1)^{\frac{55}{2}+25 k_1}\nonumber\\
&\times&\tilde{p}^{-\frac{31}{2}-\frac{25.11. (j_1-\epsilon)}{2}-75 k_1-125 j_1k_1+120\epsilon k_1}q^{-65k_1^2-\frac{25}{2}k_-\frac{15.11}{2}-\frac{25.11.k_1}{2}}\prod_{j=5j_1+1}^{5j_1+5}(1-\tilde{p}^{-j+5\epsilon})\nonumber\\
&\times&\prod_{j=0}^{\infty}\frac{1-\tilde{p}^{j-5j_1+5\epsilon}}{(1-\tilde{p}^{j-j_1+\epsilon})^5}\Big\}.
\eea
After simplification we get the expression
\bea 
&&\lim_{q\to 0}\mathcal{L}^{4d}J^{4d}=\nonumber\\
&&\Big\{\sum_{j_1=0}Q_1^{j_1}e^{-i\frac{\pi}{3}(-120a(\sigma \epsilon-\sigma j_1)^3+30b(-\sigma j_1+\sigma \epsilon)^2)}\prod_{j=0}^{\infty}(1-\tilde{p}^{j-j_1+\epsilon})^5(1-\tilde{p}^{j_1-\epsilon})^5
\prod_{j=0}^{\infty}(1-\tilde{p}^{j+j_1+1-\epsilon})^5\nonumber\\
&\times&\prod_{j=0}^{\infty}\frac{1-\tilde{p}^{j-5j_1+5\epsilon}}{(1-\tilde{p}^{j-j_1+\epsilon})^5}+\sum_{j_1=0}^{\infty}Q_1^{j_1}G_1[j_1](1-\tilde{p}^{-j_1+\epsilon})^5\prod_{j=0}^{\infty}\frac{1-\tilde{p}^{j-5j_1+5\epsilon}}{(1-\tilde{p}^{j-j_1+\epsilon})^5}\Big\}-\nonumber\\&&\Big\{\sum_{j_1=0}^{\infty}Q_1^{j_1+1}G_2[j_1]\prod_{j=5j_1+1}^{5j_1+5}(1-\tilde{p}^{-j+5\epsilon})\prod_{j=0}^{\infty}\frac{1-\tilde{p}^{j-5j_1+5\epsilon}}{(1-\tilde{p}^{j-j_1+\epsilon})^5}\Big\},\nonumber\\
\eea
where
\bea
G_1[j_1]&=&\sum_{k_1=1}^{\infty}e^{-i\frac{\pi}{3}(120a(\tau k_1)^3+30b(\tau k_1)^2)}(-1)^{5(k_1-1)}\tilde{p}^{5(k_1-1)(-j_1+\epsilon)-\frac{5}{12}-\frac{5 j_1}{2}+\frac{5\epsilon}{2}-10k_1\epsilon}q^{-5k_1^2}Q_2^{k_1},\nonumber\\
G_2[j_1]&=&\sum_{k_1=0}^{\infty}Q_2^{k_1+1}e^{-i\frac{\pi}{3}(120a(\tau (k_1+1))^3+30b(\tau (k_1+1))^2)}(-1)^{\frac{55}{2}+25 k_1}\nonumber\\
&\times&\tilde{p}^{-\frac{31}{2}-\frac{25.11. (j_1-\epsilon)}{2}-75 k_1-125 j_1k_1+120\epsilon k_1}q^{-65k_1^2-\frac{25}{2}k_-\frac{15.11}{2}-\frac{25.11.k_1}{2}}.
\eea

Altough not exactly, the last difference equation resembles schematically a certain deformation of the following 3d difference equation \cite{Jockers:2018sfl} 
\bea 
&&\mathcal{L}^{3d}J^{3d}=\nonumber\\&&\Big\{\sum_{j_1=0}Q_1^{j_1}(1-\tilde{p}^{j_1+\epsilon})^5\prod_{j=0}^{\infty}\frac{1-\tilde{p}^{j-5j_1+5\epsilon}}{(1-\tilde{p}^{j-j_1+\epsilon})^5}\Big\}-\Big\{\sum_{j_1=0}Q_1^{j_1+1}\prod_{i=5j_1+1}^{5j_1+5}(1-\tilde{p}^{i+5\epsilon})\prod_{j=0}^{\infty}\frac{1-\tilde{p}^{j-5j_1+5\epsilon}}{(1-\tilde{p}^{j-j_1+\epsilon})^5}\Big\}.\nonumber\\
\eea
\section{Second degeneration limit: $q \to 1$}\label{Sec:qto1}
A second limit of the 4d partition function that one wants to consider is $q\to 1$. In this limit the elliptic gamma function transforms as
\bea 
\lim_{q\to 1}\Gamma_e(x;p,q)=\Theta(x;p)e^{-\frac{q}{1-q}\sum_{s\ge 0}({\rm Li}_{2}(xp^s)+{\rm Li}_{2}(x^{-1}p^{s+1}))}
\eea
where $\Theta(x;p)$ is called the short Jacobi Theta function and is defined as
\bea 
\Theta(x;p)=(x;p)_{\infty}(x^{-1}p;p)_{\infty}
\eea
\subsection{$q\to 1$ limit of the 4d partition function}
The 4d J-function is given by
\bea
J^{4d}&=&\sum_{j_1\ge 0}\sum_{k_1\ge 0}Q_1^{j_1}Q_2^{k_1}e^{\frac{- i \pi }{3}(\frac{120}{\tau\sigma}(-\sigma j_1+\epsilon \sigma-\tau k_1)^3-\frac{30}{\tau\sigma}(1+\tau+\sigma)(-\sigma j_1+\epsilon \sigma-\tau k_1)^2)}\nonumber\\
&\times& \prod_{j=0}^{\infty}\prod_{k=0}^{\infty}\frac{(1-\lambda_+\tilde{P}^{j-5j_1+5\epsilon}q^{k-5k_1})}{(1-\lambda_-\tilde{P}^{j-j_1+\epsilon}q^{k-k_1})^5}\prod_{j=0}^{\infty}\prod_{k=0}^{\infty}\frac{(1-\lambda_4\tilde{P}^{j+j_1+\epsilon+1}q^{k+k_1+1})}{(1-\lambda_3\tilde{P}^{j+5j_1+5\epsilon+1}q^{k+5k_1+1})^5}.
\eea
In the $q\to 1$ limit it reduces to the following expression
\bea 
\lim_{q\to 1}J^{4d}&=&\sum_{j_1\ge 0}\sum_{k_1\ge 0}Q_1^{j_1}Q_2^{k_1}\lim_{\tau\to 0}\Big(e^{\frac{- i \pi }{3}(\frac{120}{\tau\sigma}(-\sigma j_1+\epsilon \sigma-\tau k_1)^3-\frac{30}{\tau\sigma}(1+\tau+\sigma)(-\sigma j_1+\epsilon \sigma-\tau k_1)^2)}\Big)\nonumber\\&\times&\frac{\Theta(\tilde{P}^{-5j_1+5\epsilon;\tilde{P}})^{\frac{1}{2}+5k_1}}{\Theta(\tilde{P}^{-j_1+\epsilon;\tilde{P}})^{5(\frac{1}{2}+k_1)}}e^{-\frac{1}{2}\frac{1+q}{1-q}\sum_{s\ge 0}({\rm Li}_{2}(\tilde{P}^{-5j_1+5\epsilon+s})-5{\rm Li}_{2}(\tilde{P}^{-j_1+\epsilon+s})+5{\rm Li}_{2}(\tilde{P}^{j_1+1+\epsilon+s})-{\rm Li}_{2}(\tilde{P}^{5j_1+1-5\epsilon+s}))}.\nonumber\\
\eea
\subsection{$q \to 1$ limit of the 4d difference equation }
The $q\to 1$ limit  of the 4d difference operator acting on J-function Eq. (\ref{eq:Diffeq2}) turns out to be given by the following expression
\bea
&&\Big\{\sum_{j_1\ge 0}\sum_{k_1\ge 0}Q_1^{j_1}Q_2^{k_1}\lim_{\tau\to 0}\Big(e^{\frac{- i \pi }{3}(\frac{120}{\tau\sigma}(-\sigma j_1+\epsilon \sigma-\tau k_1)^3-\frac{30}{\tau\sigma}(1+\tau+\sigma)(-\sigma j_1+\epsilon \sigma-\tau k_1)^2)}\Big)\nonumber\\&&e^{\frac{1}{2}\frac{1+q}{1-q}( \sum_{j=-5}^{-1}{\rm Li}_2(\tilde{P}^{j-5j_1+5\epsilon})+\sum_{j=0}^{4}{\rm Li}_2(\tilde{P}^{j+5j_1+1-5\epsilon}))}\nonumber\\&&\times \frac{\prod_{j=-5}^{-1}(1-\tilde{P}^{j-5 j_1+5\epsilon})^{\frac{1}{2}+5 k_1}}{\prod_{j=0}^{4}(1-\tilde{P}^{j+5 j_1+1-5\epsilon})^{\frac{1}{2}+5 k_1}}\frac{\prod_{j=-5}^{-1}(1-\tilde{P}^{j-5 j_1+5\epsilon})^{5}}{\prod_{j=0}^{4}(1-\tilde{P}^{j+5 j_1+1-5\epsilon})^{5}}\Theta(\tilde{P}^{-5j_1+5\epsilon};\tilde{P})^5\nonumber\\&&-\nonumber\\&&\sum_{j_1\ge 0}\sum_{k_1\ge 0}Q_1^{j_1+1}Q_2^{k_1+1}\lim_{\tau\to 0}\Big(e^{\frac{- i \pi }{3}(\frac{120}{\tau\sigma}(-\sigma (j_1+1)+\epsilon \sigma-\tau (k_1+1))^3-\frac{30}{\tau\sigma}(1+\tau+\sigma)(-\sigma (j_1+1)+\epsilon \sigma-\tau (k_1+1))^2)}\Big)\nonumber\\&& e^{-\frac{5}{2}\frac{1+q}{1-q}( {\rm Li}_2(\tilde{P}^{j_1-\epsilon+s})+ {\rm Li}_2(\tilde{P}^{-j_1+\epsilon}))}\Theta(\tilde{P}^{-j_1+\epsilon};\tilde{P})^5\Big\}\nonumber\\&&e^{-\frac{1}{2}\frac{1+q}{1-q}\Big(\sum_{s\ge 0}( {\rm Li}_2(\tilde{P}^{-5j_1+5\epsilon+s})-5 {\rm Li}_2(\tilde{P}^{-j_1+\epsilon+s})+5 {\rm Li}_2(\tilde{P}^{j_1+1-\epsilon+s})- {\rm Li}_2(\tilde{P}^{5j_1+1-5\epsilon+s}))\Big)}\nonumber\\&&
\frac{\Theta(\tilde{P}^{-5j_1+5\epsilon};\tilde{P})^{\frac{1}{2}+5k_1}}{\Theta(\tilde{P}^{-j_1+\epsilon};\tilde{P})^{5(\frac{1}{2}+k_1)}}=0.
\eea
\section{4d generalization of 3d permutation-equivariant quantum K-theory invariants}\label{Sec:4dgenQK}
In permutation-equivariant quantum K-theory the Gromov-Witten invariants are defined as the multiplicities of the representations of the group of permutations that act on the marked points in 
 the moduli spaces of stable maps. In the construction of quantum K-theory invariants, Givental's reconstruction theorems play a central role. To explain the idea behind it, we have to first define a few terms. We define $\mathcal{K}$ as $\mathcal{K}=\mathcal{K}_+\oplus \mathcal{K}_-$. A subset $\mathcal{L}\subset \mathcal{K}$ is called the range of the $J$-function if it satisfies certain conditions.
 A function $g\in \mathcal{K}$ lies in the range $\mathcal{L}$ if and only if its Laurent expansion near $q=\xi$ satisfies the following three conditions:\\
 1) $g_{\xi=1}=(1-q)e^{\frac{\tau}{1-q}}\times$ power series in $(q-1)$,\\
 2) For $\xi \ne 1$ a primitive n-th root of unity $g_{\xi}(q^{1/n}/\xi)=\Psi^n(g_1/(1-q))\times$ power series in $(q-1)$, where $\Psi^n$ is the Adams operation defined on $q$ by $\Psi^n(q)=q^n$.\\
 3) For $\xi\ne 0,\infty$ not a root of unity, $g_{\xi}(q/\xi)$ is a power series in $(q-1)$.\\
 A useful property of $\mathcal{L}$ is that it is invariant under the action of a large group of symmetries generated by certain difference operators. The essence of the reconstruction theorem of Givental is that we can construct the whole $\mathcal{L}$ starting from one point on $\mathcal{L}$ if symmetries are properly utilized.

The 4d equivalent of the J-function of quantum K-theory with non-zero input ($t\ne 0$) is given by

\bea\label{eq:J4d}
J(t)&=&\sum_{j_1=0}^{\infty}\sum_{k_1=0}^{\infty}Q_1^{j_1}Q_2^{k_1}e^{-i\frac{\pi}{3}(120a(-\sigma j_1+\epsilon\sigma-\tau k_1+\eta\tau)^3+20b(-\sigma j_1+\epsilon\sigma-\tau k_1+\eta\tau)^2)}\nonumber\\&\times&
\prod_{j=0}^{\infty}\prod_{k=0}^{\infty}\frac{1-\tilde{P}^{j-5j_1+5\epsilon}q^{k-5k_1}}{(1-\tilde{P}^{j-j_1+\epsilon}q^{k-k_1})^5}\prod_{j=0}^{\infty}\prod_{k=0}^{\infty}\frac{(1-\tilde{P}^{j+j_1+1-\epsilon}q^{k+k_1+1})^5}{1-\tilde{P}^{j+5j_1+1-5\epsilon}q^{k+5k_1+1}}.
\eea

Note that as defined in \ref{eq:PH}, $P=\tilde{P}^{-\epsilon}$. Below we will redefine $P$ as $P=1-H$. Following Givental \cite{Givental2015PermutationequivariantQK1,Givental2015PermutationequivariantQK2,Givental2015PermutationequivariantQK3,Givental2015PermutationequivariantQK4,Givental2015PermutationequivariantQK5,Givental2015PermutationequivariantQK6,Givental2015PermutationequivariantQK7,Givental2015PermutationequivariantQK8,Givental2015ExplicitRI}, we can apply to $J(t)$ a 4d generalization of the Givental transformation 

\bea\label{eq:Givental4d}
e^{\sum_{r=1}^{\infty}\frac{\sum_{l=0}^3\Psi_r(\epsilon_l)(1-H)^{lr}\tilde{P}^{ l r Q_1\partial_{Q_1}}(1+q^{lrQ_2\partial_{Q_2}})}{r(1-\tilde{P}^r)(1-q^r)}},
\eea
to get $J(0)$.\\The process of obtaining 4d invariants consists of two steps,\\
1) Apply the generalized Givental transformation Eq. (\ref{eq:Givental4d}) to $J(t)$ to get $J(0)$, by properly choosing the input paramters .\\
2) Apply another Givental's transformation to $J(0)$ to get 4d invariants.\\

We now consider the following generalization of Givental's reconstruction theorems\\
\textbf{Proposition}:\\
Let $I=\sum_{j_1,k_1} I_{j_1,k_1}Q^{j_1}Q^{k_1}$ be a point in the range of $\mathcal{L}\subset \mathcal{K}$ of the $J$-function of the generalized permutation-equivariant quantum K-theory on $X$. Then the following family also lies in $\mathcal{L}$
\bea\label{eq:proposition}
\sum_{j_1,k_1} I_{j_1,k_1}Q^{j_1}Q^{k_1}e^{\sum_{r=1}^{\infty}\frac{\sum_{l=0}^3\Psi_r(\epsilon_l)(1-H)^{lr}\tilde{P}^{ l r j_1}(1+q^{lrk_1})}{r(1-\tilde{P}^r)(1-q^r)}}.
\eea
Notice that the generalized Givental transformation given in Eq. (\ref{eq:Givental4d}) reduces to the Givental transformation given in \cite{Jockers:2018sfl} in the $q\to 0$ limit

\bea\label{eq:Givental3d}
e^{\sum_{r=1}^{\infty}\frac{\sum_{l=0}^3\Psi_r(\epsilon_l)(1-H)^{lr}\tilde{P}^{ l r Q_1\partial_{Q_1}}}{r(1-\tilde{P}^r)}}.
\eea
We will prove and investigate the implications of this proposition in a later work.

\subsection{$q\to 0$ limit }
Consider the generalized Givental transformation in Eq. (\ref{eq:Givental4d}) acting on $J(t)$  given in Eq. (\ref{eq:J4d})
\bea
&&J(t^{\prime}):=(1-q)(1-\tilde{P}^{-1})e^{\sum_{r=1}^{\infty}\frac{\sum_{l=0}^3\Psi_r(\epsilon_l)(1-H)^{lr}\tilde{P}^{ l r Q_1\partial_{Q_1}}(1+q^{lrQ_2\partial_{Q_2}})}{r(1-\tilde{P}^r)(1-q^r)}}J(t)\nonumber\\&&=(1-q)(1-\tilde{P}^{-1})\sum_{j_1=0}^{\infty}\sum_{k_1=0}^{\infty}
e^{\sum_{r=1}^{\infty}\frac{\sum_{l=0}^3\Psi_r(\epsilon_l)(1-H)^{lr}\tilde{P}^{ l r j_1}(1+q^{lrk_1})}{r(1-\tilde{P}^r)(1-q^r)}}
Q_1^{j_1}Q_2^{k_1}e^{F_1[j_1,k_1]}\nonumber\\&\times&
\prod_{j=0}^{\infty}\prod_{k=0}^{\infty}\frac{1-\tilde{P}^{j-5j_1+5\epsilon}q^{k-5k_1}}{(1-\tilde{P}^{j-j_1+\epsilon}q^{k-k_1})^5}\prod_{j=0}^{\infty}\prod_{k=0}^{\infty}\frac{(1-\tilde{P}^{j+j_1+1-\epsilon}q^{k+k_1+1})^5}{1-\tilde{P}^{j+5j_1+1-5\epsilon}q^{k+5k_1+1}},
\eea
where $e^{F_1[j_1,k_1]}= e^{-i\frac{\pi}{3}(120a(-\sigma j_1+\epsilon\sigma-\tau k_1+\eta\tau)^3+20b(-\sigma j_1+\epsilon\sigma-\tau k_1+\eta\tau)^2)}$.\\ In the $q\to 0$ limit we get the following transformed $J$-function
\bea 
J(t^{\prime})&=&(1-\tilde{P}^{-1})+(1-\tilde{P}^{-1})(q_{10}+q_{11}H++q_{12}H^2++q_{13}H^3)Q_1\nonumber\\&+&(1-\tilde{P}^{-1})(q_{20}+q_{21}H++q_{22}H^2++q_{23}H^3)Q_2+...
\eea
With the following definition of $\epsilon[i,j]$ functions 
\bea 
\epsilon[1,0]&=&\epsilon_{1100} + Q_1 \epsilon_{1101} + Q_2 \epsilon_{2101}+...,\nonumber\\
\epsilon[1,1]&=&\epsilon_{1110} + Q_1 \epsilon_{1111} + Q_2 \epsilon_{2111}+...,\nonumber\\
\epsilon[1,2]&=&\epsilon_{1120} + Q_1 \epsilon_{1121} + Q_2 \epsilon_{2121}+...,\nonumber\\
\epsilon[1,3]&=&\epsilon_{1130} + Q_1 \epsilon_{1131} + Q_2 \epsilon_{2131}+....\nonumber\\
\eea

the coefficients $q_{ij}$ are given below
\bea\label{eq:6.8}
q_{10}&=&\frac{(\tilde{P}+1)^2 \left(\tilde{P}^2+1\right) \left(\tilde{P}^2+\tilde{P}+1\right)
   \left(\tilde{P}^4+\tilde{P}^3+\tilde{P}^2+\tilde{P}+1\right) e^{F_1[1,0]-F_1[0,0]}}{\tilde{P}^{10}}\nonumber\\&+&\frac{\epsilon_{1101}+\epsilon_{1111}+\epsilon_{1121}+\epsilon_{1131}}{1-\tilde{P}},\nonumber\\
q_{11}&=&\frac{5 \left(\tilde{P}^9+4 \tilde{P}^8+10 \tilde{P}^7+18 \tilde{P}^6+26 \tilde{P}^5+30 \tilde{P}^4+28 \tilde{P}^3+21 \tilde{P}^2+12
   \tilde{P}+4\right) e^{F_1[1,0]-F_1[0,0]}}{(\tilde{P}-1) \tilde{P}^{10}}\nonumber\\&+&\frac{\epsilon_{1111}+2 \epsilon_{1121}+3
  \epsilon_{1131}}{\tilde{P}-1},\nonumber\\
  q_{12}&=&-\frac{5 \left(2 \tilde{P}^{11}+10 \tilde{P}^{10}+24 \tilde{P}^9+42 \tilde{P}^8+51 \tilde{P}^7+45 \tilde{P}^6+13 \tilde{P}^5-34 \tilde{P}^4-71
   \tilde{P}^3-85 \tilde{P}^2-74 \tilde{P}-38\right) }{(\tilde{P}-1)^2 \tilde{P}^{10}}\nonumber\\&&e^{F_1[1,0]-F_1[0,0]}-\frac{\epsilon_{1121}+3 \epsilon_{1131}}{\tilde{P}-1},\nonumber\\
   q_{13}&=&\frac{5 \left(2 \tilde{P}^{12}+16 \tilde{P}^{11}+36 \tilde{P}^{10}+50 \tilde{P}^9+14 \tilde{P}^8-56 \tilde{P}^7-176 \tilde{P}^6-262
   \tilde{P}^5-217 \tilde{P}^4-102 \tilde{P}^3+43 \tilde{P}^2\right) }{(\tilde{P}-1)^3
   \tilde{P}^{10}}\nonumber\\&&e^{F_1[1,0]-F_1[0,0]}+\frac{194 \tilde{P}+228}{(\tilde{P}-1)^3
   \tilde{P}^{10}}e^{F_1[1,0]-F_1[0,0]}+\frac{\epsilon_{1131}}{(\tilde{P}-1)^3}.
\eea
Note the ubiquitous phase factors $e^{F_1[1,0]-F_1[0,0]}$. More precisely, they should be written as $\lim_{\tau\to \infty}e^{F_1[1,0]-F_1[0,0]}$.  
\bea \label{eq:6.11}
q_{20}&=&\frac{\left(\left(10 \tilde{P}^2+15 \tilde{P}+9\right) q^2+\left(20 \tilde{P}^3+36 \tilde{P}^2+31 \tilde{P}+15\right) q^3\right)
   e^{F_1[0,1]-F_1[0,0]}}{q^{10}}\nonumber\\&+&\frac{\epsilon_{2101}+\epsilon_{2111}+\epsilon_{2121}+\epsilon_{2131}}{1-\tilde{Pt}}\nonumber\\&+&\frac{\left(\left(35
   \tilde{P}^4+70 \tilde{P}^3+70 \tilde{P}^2+46 \tilde{P}+20\right) q^4+\left(56 \tilde{P}^5+120 \tilde{P}^4+130 \tilde{P}^3+95 \tilde{P}^2+51
   \tilde{P}+22\right) q^5\right)
  }{q^{10}}\nonumber\\&\times&e^{F_1[0,1]-F_1[0,0]}\nonumber\\&+&\frac{\left(\left(84 \tilde{P}^6+189 \tilde{P}^5+215 \tilde{P}^4+164 \tilde{P}^3+89 \tilde{P}^2+40 \tilde{P}+20\right)
   q^6+\left(120 \tilde{P}^7+280 \tilde{P}^6+329 \tilde{P}^5\right)\right)
  }{q^{10}}\nonumber\\&& e^{F_1[0,1]-F_1[0,0]}\nonumber\\&+&\frac{\left(255 \tilde{P}^4+130 \tilde{P}^3+43 \tilde{P}^2+14 \tilde{P}+15\right)
   q^7+\left(165 \tilde{P}^8+396 \tilde{P}^7+476 \tilde{P}^6+370 \tilde{P}^5\right)}{q^{10}}\nonumber\\&& e^{F_1[0,1]-F_1[0,0]}\nonumber\\&+&\frac{+\left(170 \tilde{P}^4+20 \tilde{P}^3-31 \tilde{P}^2-17
   \tilde{P}+9\right) q^8+\left(220 \tilde{P}^9+540 \tilde{P}^8+660 \tilde{P}^7+511 \tilde{P}^6+205 \tilde{P}^5\right)
   }{q^{10}}\nonumber\\&& e^{F_1[0,1]-F_1[0,0]}\nonumber\\&+&\frac{-\left(40 \tilde{P}^4-129
   \tilde{P}^3-100 \tilde{P}^2-41 \tilde{P}+4 q^9\right)+\left(286 \tilde{P}^{10}+715 \tilde{P}^9+885 \tilde{P}^8+680 \tilde{P}^7\right)
   }{q^{10}}\nonumber\\&& e^{F_1[0,1]-F_1[0,0]}\nonumber\\&+&\frac{\left(231
   \tilde{P}^6-148 \tilde{P}^5-294 \tilde{P}^4-243 \tilde{P}^3-133 \tilde{P}^2-50 \tilde{P}+1\right) q^{10}+4 (\tilde{P}+1) q+1
    }{q^{10}}\nonumber\\&&e^{F_1[0,1]-F_1[0,0]}
\eea
\bea \label{eq:6.12}
q_{21}&=&\frac{1}{q^{10}}\Big(\frac{q^{10} (\epsilon_{2111}+2 \epsilon_{2121}+3 \epsilon_{2131})}{\tilde{P}-1}-5( \left(40 \text{P}^2+59
   \tilde{P}+37\right) q^2+80 \tilde{P}^3 q^3\nonumber\\&+&\left(140 \tilde{P}^2+123 \tilde{P}+65\right) q^3+\left(140 \tilde{P}^4+270 \tilde{P}^3+270
   \tilde{P}^2+189 \tilde{P}+95\right) q^4\nonumber\\&+&\left(224 \tilde{P}^5+460 \tilde{P}^4+490 \tilde{P}^3+372 \tilde{P}^2+229 \tilde{P}+121\right)
   q^5\nonumber\\&+&\left(336 \tilde{P}^6+721 \tilde{P}^5+795 \tilde{P}^4+614 \tilde{P}^3+376 \tilde{P}^2+223 \tilde{P}+139\right) q^6\nonumber\\&+&\left(480
   \tilde{P}^7+1064 \tilde{P}^6+1197 \tilde{P}^5+915 \tilde{P}^4+510 \tilde{P}^3+257 \tilde{P}^2+172 \tilde{P}+149\right) q^7\nonumber\\&+&\left(660
   \tilde{P}^8+1500 \tilde{P}^7+1708 \tilde{P}^6+1275 \tilde{P}^5+605 \tilde{P}^4+174 \tilde{P}^3+58 \tilde{P}^2+99 \tilde{P}+153\right)
   q^8\nonumber\\&+&\left(880 \tilde{P}^9+2040 \tilde{P}^8+2340 \tilde{P}^7+1694 \tilde{P}^6+635 \tilde{P}^5-75 \tilde{P}^4-251 \tilde{P}^3-128
   \tilde{P}^2+31 \tilde{P}+154\right) \nonumber\\&q^9&+\left(1144 \tilde{P}^{10}+2695 \tilde{P}^9+3105 \tilde{P}^8+2172 \tilde{P}^7+574 \tilde{P}^6-539
   \tilde{P}^5-803 \tilde{P}^4-545 \tilde{P}^3\right) q^{10}\nonumber\\&-&225\left(\left(\left( \tilde{P}^2-12 \tilde{P}+154\right) q^{10}+16 (\tilde{P}+1) q+4\right)\right)
   e^{F_1[0,1]-F_1[0,0]}\Big)
\eea
\bea\label{eq:6.13}
q_{22}&=&\frac{1}{q^10} \Big(5e^{F_1[0,1]-F_1[0,0]} \left(\left(718 \tilde{P}^3+1214 \tilde{P}^2+1072 \tilde{P}+615\right) q^3+\left(1218 \tilde{P}^4+2244 \tilde{P}^3\right) q^4\right) \nonumber\\&+&\left(2180 \tilde{P}^2+1582 \tilde{P}+917\right)
   q^4+\left(1876 \tilde{P}^5+3644 \tilde{P}^4+3642 \tilde{P}^3+2693 \tilde{P}^2\right)q^5\nonumber\\&+&\left(1810 \tilde{P}+1206\right) q^5+\left(2688 \tilde{P}^6+5404 \tilde{P}^5+5387 \tilde{P}^4+3704 \tilde{P}^3+2149 \tilde{P}^2\right)q^6\nonumber\\&+&\left(1590 \tilde{P}+1450\right)
   q^6+\left(3636\tilde{P}^7+7476 \tilde{P}^6+7294 \tilde{P}^5+4331 \tilde{P}^4\right)q^7\nonumber\\&+&\left(1390 \tilde{P}^3+450 \tilde{P}^2+953 \tilde{P}+1643\right) q^7+\left(4686 \tilde{P}^8+9768 \tilde{P}^7+9184 \tilde{P}^6\right) q^8\nonumber\\&+&\left(4246 \tilde{P}^5-1017
   \tilde{P}^4-2886 \tilde{P}^3-1859 \tilde{P}^2+100 \tilde{P}+1794\right) q^8\nonumber\\&+&\left(5786 \tilde{P}^9+12138 \tilde{P}^8+10812 \tilde{P}^7+3073 \tilde{P}^6-5620 \tilde{P}^5-9038 \tilde{P}^4-7395 \tilde{P}^3\right)q^9\nonumber\\&-&\left(3888 \tilde{P}^2-732
   \tilde{P}+1921\right) q^9+\left(6864 \tilde{P}^{10}+14388 \tilde{P}^9+11859 \tilde{P}^8+384 \tilde{P}^7\right)q^{10}\nonumber\\&-&\left(12957 \tilde{P}^6-18552 \tilde{P}^5-15958 P^4-10071 \tilde{P}^3-4969 \tilde{P}^2-1359 \tilde{P}+2038\right)
   q^{10}\nonumber\\&+&q^2 \left. \left(532 \tilde{P}+368 \tilde{P}^2+347\right)+150 (\tilde{P}+1) q+38\right)+\frac{q^{10} (\epsilon_{2121}+3 \epsilon_{2131})}{1-\tilde{P}}\Big)\nonumber
\eea
\bea \label{eq:6.14}
q_{23}&=&\frac{1}{q^{10}}\Big(5 (-\left(2020 \tilde{P}^2+2884 \tilde{P}+1993\right) q^2-\left(3622 \tilde{P}^3+5952 \tilde{P}^2\right)q^3\nonumber\\&+&\left(5328 \tilde{P}+3471\right)
   q^3-\left(5404 \tilde{P}^4+9528 \tilde{P}^3+8766 \tilde{P}^2+6832 \tilde{P}+5095\right) q^4\nonumber\\&-&\left(6804 \tilde{P}^5+12374 \tilde{P}^4+10248
   \tilde{P}^3+6572 \tilde{P}^2+5792 \tilde{P}+6603\right) q^5\nonumber\\&-&\left(6888 \tilde{P}^6+12348 \tilde{P}^5+6631 \tilde{P}^4-1694
   \tilde{P}^3-3968 \tilde{P}^2+1350 \tilde{P}+7819\right) q^6\nonumber\\&+&\left(-4260 \tilde{P}^7-6160 \tilde{P}^6+6608 \text{Pt}^5+23373
   \tilde{P}^4+30636 \tilde{P}^3+22034 \tilde{P}^2+6053 \tilde{P}-8687\right) q^7\nonumber\\&+&\left(3036 \tilde{P}^8+10896 \tilde{P}^7+35700
   \tilde{P}^6+65058 \tilde{P}^5+80843 \tilde{P}^4+69258 \tilde{P}^3+42724 \tilde{P}^2\right)q^8\nonumber\\&+&\left( 15054 \tilde{P}-9221\right) q^8+\left(17622
   \tilde{P}^9+45234 \tilde{P}^8+88944 \tilde{P}^7+134708 \tilde{P}^6\right)\nonumber\\&+&\left(161156 \tilde{P}^5+144833 \tilde{P}^4+103015 \tilde{P}^3+59754
   \tilde{P}^2+24142 \tilde{P}-9471\right) q^9\nonumber\\&+&\left(42900 \tilde{P}^{10}+105292 \tilde{P}^9+177099 \tilde{P}^8+241844 \tilde{P}^7+277802
   \tilde{P}^6+251352 \tilde{P}^5\right)q^{10}\nonumber\\&+&\left(185331 \tilde{P}^4+118297 \text{Pt}^3+69092 \text{Pt}^2+32211 \tilde{P}-9473\right) q^{10}-2 (435
   \tilde{P}+439) q-228) \nonumber\\&&e^{F_1[0,1]-F_1[0,0]}+\frac{q^{10} \epsilon_{2131}}{\tilde{P}-1}\Big).
\eea
It is important to remember that the input parameters, {\it i.e.}, the epsilon functions, depend on variables $\tilde{P},q$. So in taking the $q\to 0$ limit, their proper Laurent series expansions are considered.
\section*{Analysis of the coefficients $q_{10},q_{11},q_{12},q_{13}$ of the $Q_1$ term}
The defining property of the input parameters $\epsilon_{1101} , \epsilon_{1111} ,\epsilon_{1121} , \epsilon_{1131}$ is that they take values in $\mathbb{C}[\tilde{P}^{-1},\tilde{P},q,q^{-1}]$. The remaining piece of the J-function is described by rational functions and it has the following properties
\\
a) It vanishes at $q=\infty,\tilde{P}^{-1}=\infty$\\
b) It vanishes at $q=\infty,\tilde{P}^{-1}=0$\\
c) It vanishes at $q=0,\tilde{P}^{-1}=\infty$\\
d) It does not have a pole at $q=0,\tilde{P}^{-1}=0$.\\
The correlation functions belong to the latter class of rational functions.\\
In the $q\to 0$ limit, the idea is to project the rational functions of $\tilde{P}$ in each of $q_{10},q_{11},q_{12},q_{13}$ into their polar and non-polar parts. The projection is performed by polynomial division. The remainder is the contribution at that order in Novikov variables to the correlation function, whereas quotient is canceled by properly choosing the input parameters. Following this logic we get the following result
\bea 
&&\epsilon_{1131}=5 (228 \tilde{P}^{-11}+650 \tilde{P}^{-10}+1115 \tilde{P}^{-9}+1478 \tilde{P}^{-8}+1624 \tilde{P}^{-7}+1508 \tilde{P}^{-6}+1216 \tilde{P}^{-5}+\nonumber\\&&\qquad 868 \tilde{P}^{-4}+534 \tilde{P}^{-3}+250 \tilde{P}^{-2}+2 \tilde{P}^{-1}-230)e^{-F_1[0, 0] + F_1[1, 0]}\nonumber\\
&&\epsilon_{1121} + 3 \epsilon_{1131}=-5 (38 \tilde{P}^{-11}+112 \tilde{P}^{-10}+197 \tilde{P}^{-9}+268 \tilde{P}^{-8}+302 \tilde{P}^{-7}+289 \tilde{P}^{-6}+244 \tilde{P}^{-5}+\nonumber\\&&\qquad193 \tilde{P}^{-4}+151 \tilde{P}^{-3}+127 \tilde{P}^{-2}+117 \tilde{P}^{-1}+115)
   e^{F_1[1,0]-F_1[0,0]}\nonumber\\
&&\epsilon_{1111} + 2 \epsilon_{1121} + 3 \epsilon_{1131}=5 \tilde{P}^{-2} (4 \tilde{P}^{-9}+12 \tilde{P}^{-8}+21 \tilde{P}^{-7}+28 \tilde{P}^{-6}+30 \tilde{P}^{-5}+26 \tilde{P}^{-4}+18 \tilde{P}^{-3}+\nonumber\\&& \qquad 10 \tilde{P}^{-2}+4 \tilde{P}^{-1}+1) e^{F_1[1,0]-F_1[0,0]}\nonumber\\
&&\epsilon_{1101} + \epsilon_{1111} + \epsilon_{1121} + \epsilon_{1131}=(1-\tilde{P}^{-1}) (\tilde{P}^{-1}+1)^2 \left(\tilde{P}^{-2}+1\right)\left(\tilde{P}^{-2}+\tilde{P}^{-1}+1\right) \nonumber\\&& \qquad\qquad\qquad\left(\tilde{P}^{-4}+\tilde{P}^{-3}+\tilde{P}^{-2}+\tilde{P}^{-1}+1\right)  e^{F_1[1,0]-F_1[0,0]}.
\eea
Note that the expressions on the right hand side exactly match with the input $t^{sym}$ given in Equation (6.35) of \cite{Jockers:2018sfl} if we make the identification $q=\tilde{P}^{-1}$
\bea 
t^{sym}&=&\Big((1-q) (q+1)^2 \left(q^{2}+1\right)\left(q^{2}+q^{1}+1\right)(q^{4}+q^{3}+q^{2}+q^{1}+1)\Phi_0\nonumber\\&+& 5 q^{2} (4 q^{9}+12 q^{8}+21 q^{7}+28 q^{6}+30 q^{5}+26 q^{4}+18 q^{3}+10 q^{2}+4 q+1)\Phi_1\nonumber\\&+&5(115+117 q+...+112 q^10+38 q^11)\Phi_2\nonumber\\&+&2(-230+2q+...+228 q^11)\Phi_3\Big)Q+.....
\eea
In the polynomial division of the rational functions in $q_{10},q_{11},q_{12},q_{13}$, the remainders are the polar parts
given below
\bea 
&&J^{polar}=(1-\tilde{P}^{-1})+\Big( \frac{575 }{1-\tilde{P}^{-1}}\Phi_2+1150 \frac{1-2\tilde{P}^{-1}}{(1-\tilde{P}^{-1})^2}\Phi_3\Big)Q_1+\ldots
\eea
Notice that it exactly matches with the function $J^{3d}(0)$ given in Equation (6.38) of \cite{Jockers:2018sfl} up to order $Q_1=Q$.

\section*{Analysis of the coefficients $q_{20},q_{21},q_{22},q_{23}$ of the $Q_2$ term}
We know that the input parameters $\epsilon_{2101},\epsilon_{2111},\epsilon_{2121},\epsilon_{2131}$ take values in the ring of Laurent polynomials $\mathbb{C}[\tilde{P}^{-1},\tilde{P},q,q^{-1}]$. If we observe the expressions for $q_{20},q_{21},q_{22},q_{23}$ in Eqs. (\ref{eq:6.11},\ref{eq:6.12},\ref{eq:6.13},\ref{eq:6.14}), we notice that there is a pole $\frac{1}{q^{10}}$ at $q=0$ in each of them. Due to poles, these terms are the elements of the ring of Laurent polynomials $\mathbb{C}[\tilde{P}^{-1},\tilde{P},q,q^{-1}]$ and so, can be absorbed in the choice of $\epsilon_{2101},\epsilon_{2111},\epsilon_{2121},\epsilon_{2131}$. Therefore, in the limit $q\to 0$ there is no contribution to $J^{3d}(0)$ at order $Q_2$.\\ Thus upto order $Q_1$ and $Q_2$ we get in the $q\to 0$ limit 
\bea 
&&J^{3d}_{q=0}(0)=(1-\tilde{P}^{-1})+\Big( \frac{575 }{1-\tilde{P}^{-1}}\Phi_2+1150 \frac{1-2\tilde{P}^{-1}}{(1-\tilde{P}^{-1})^2}\Phi_3\Big)Q_1+\ldots
\eea
A similar analysis shows that to next higher order, $Q_2^2$ and $Q_1Q_2$ terms do not contribute in the $q\to 0$ limit and we obtain
\bea 
&&J^{3d}_{q=0}(0)=(1-\tilde{P})+\Big( \frac{575 }{1-\tilde{P}}\Phi_2+1150 \frac{1-2\tilde{P}}{(1-\tilde{P})^2}\Phi_3\Big)Q_1+\nonumber\\&&\Big( \frac{25(9794+19496 \tilde{P}+9725\tilde{P}^2)\Phi_2}{(1-\tilde{P})(1+\tilde{P})^2}+\frac{50(7380+9748\tilde{P}-14760\tilde{P}^2-29244\tilde{P}^3-12139\tilde{P}^4)\Phi_3}{(1-\tilde{P})^2(1+\tilde{P})^3}\Big)Q_1^2\nonumber\\&&+ \ldots
\eea

which matches the 3d J-function $J^{3d}_{q=0}(0)$ given in Equation (6.38) of \cite{Jockers:2018sfl}.
\subsection{$q\to 1$ limit}
 As in the last section, the transformed J-function is given by
\bea
&&J(t^{\prime}):=(1-q)(1-\tilde{P}^{-1})e^{\sum_{r=1}^{\infty}\frac{\sum_{l=0}^3\Psi_r(\epsilon_l)(1-H)^{lr}\tilde{P}^{ l r Q_1\partial_{Q_1}}(1+q^{lrQ_2\partial_{Q_2}})}{r(1-\tilde{P}^r)(1-q^r)}}J(t)\nonumber\\&&=(1-q)(1-\tilde{P}^{-1})\sum_{j_1=0}^{\infty}\sum_{k_1=0}^{\infty}
e^{\sum_{r=1}^{\infty}\frac{\sum_{l=0}^3\Psi_r(\epsilon_l)(1-H)^{lr}\tilde{P}^{ l r j_1}(1+q^{lrk_1})}{r(1-\tilde{P}^r)(1-q^r)}}
Q_1^{j_1}Q_2^{k_1}e^{F_1[j_1,k_1]}\nonumber\\&\times&
\prod_{j=0}^{\infty}\prod_{k=0}^{\infty}\frac{1-\tilde{P}^{j-5j_1+5\epsilon}q^{k-5k_1}}{(1-\tilde{P}^{j-j_1+\epsilon}q^{k-k_1})^5}\prod_{j=0}^{\infty}\prod_{k=0}^{\infty}\frac{(1-\tilde{P}^{j+j_1+1-\epsilon}q^{k+k_1+1})^5}{1-\tilde{P}^{j+5j_1+1-5\epsilon}q^{k+5k_1+1}},
\eea
where $e^{F_1[j_1,k_1]}= e^{-i\frac{\pi}{3}(120a(-\sigma j_1+\epsilon\sigma-\tau k_1+\eta\tau)^3+20b(-\sigma j_1+\epsilon\sigma-\tau k_1+\eta\tau)^2)}$.\\ In the $q\to 1$ limit we get the following transformed $J$-function
\bea 
J(t^{\prime})&=&(1-\tilde{P}^{-1})(q_{10}+q_{11}H++q_{12}H^2++q_{13}H^3)Q_1\nonumber\\&+&(1-\tilde{P}^{-1})(q_{20}+q_{21}H++q_{22}H^2++q_{23}H^3)Q_2+...
\eea
With the following definition of $\epsilon[i,j]$ functions 
\bea 
\epsilon[1,0]&=&\epsilon_{1100} + Q_1 \epsilon_{1101} + Q_2 \epsilon_{2101}+...,\nonumber\\
\epsilon[1,1]&=&\epsilon_{1110} + Q_1 \epsilon_{1111} + Q_2 \epsilon_{2111}+...,\nonumber\\
\epsilon[1,2]&=&\epsilon_{1120} + Q_1 \epsilon_{1121} + Q_2 \epsilon_{2121}+...,\nonumber\\
\epsilon[1,3]&=&\epsilon_{1130} + Q_1 \epsilon_{1131} + Q_2 \epsilon_{2131}+....\nonumber\\
\eea

the coefficients $q_{ij}$ are given below. Now in the $q\to 1$ limit the expressions are given by
\bea 
q_{10}&=&\frac{-2}{\tilde{P}} (\epsilon_{1101}+\epsilon_{1111}+\epsilon_{1121}+\epsilon_{1131}),\nonumber\\
q_{11}&=&\frac{2}{\tilde{P}} (\epsilon_{1111}+2 \epsilon_{1121}+3 \epsilon_{1131})\nonumber\\
q_{12}&=&-\frac{2}{\tilde{P}} ( \epsilon_{1121}+3 \epsilon_{1131})\nonumber\\
q_{13}&=&\frac{2}{\tilde{P}}\epsilon_{1131}
\eea
and 
\bea 
q_{20}&=&-\frac{2 (\epsilon_{2101}+\epsilon_{2111}+\epsilon_{2121}+\epsilon_{2131})}{\tilde{P}}\nonumber\\
q_{21}&=&\frac{2 \left(-385 (\tilde{P}-1) e^{F_1[0,1]-F_1[0,0]}+\epsilon_{2111}+2 \epsilon_{2121}+3
 \epsilon_{2131}\right)}{\tilde{P}}\nonumber\\
q_{22}&=&\frac{\frac{5 (\tilde{P} (3853 q-3968)-1923 q+2038) e^{F_1[0,1]-F_1[0,0]}}{q-1}-2 \epsilon_{2121}-6 \epsilon_{2131}}{\tilde{P}}\nonumber\\
q_{23}&=&\frac{1}{\tilde{P}}(-\frac{5 \left(41543 \tilde{P}^2-50156 \tilde{P}+9473\right) e^{F_1[0,1]-F_1[0,0]}}{\tilde{P}-1}+\frac{1150 (21 \tilde{P}-1)
   e^{F_1[0,1]-F_1[0,0]}}{q-1}\nonumber\\&+&\frac{1150 (\tilde{P}-1) e^{F_1[0,1]-F_1[0,0]}}{(q-1)^2}+2 \epsilon_{2131})
\eea
The first fact to notice in the limit $q\to 1$ is that the roles of $Q_1$ and $Q_2$ are interchanged.
It is obvious that the only option for $q_{10},q_{11},q_{12},q_{13}$ is
\bea 
\epsilon_{1131}=0,\quad \epsilon_{1121}=0,\quad \epsilon_{1111}=0,\quad \epsilon_{1101}=0.
\eea
Therefore, only the coefficients of $Q_2$ contribute non-trivially to the J-function $J^{3d}(0)$ in the $q\to 1$ limit.
\bea 
J^{3d}_{q=1}(t=0)&=&(1-\tilde{P}^{-1})(q_{20}+q_{21}H++q_{22}H^2++q_{23}H^3)Q_2+...\nonumber\\
\eea
\subsection{A few comments on $SL(2,\mathbb{Z})$ action}
Note that we obtain $J^{3d}_{q=0}$ by taking $q\to 0$ limit of $J^{4d}$ and $J^{3d}_{q=1}$ by taking $q\to 1$ limit of $J^{4d}$. This means that there will be two types of 4d invariants that we can obtain by acting with the transformation \ref{eq:Givental4d} on $J^{3d}_{q=0}$ and $J^{3d}_{q=1}$. There is a possibility that these two invariants are connected by certain $SL(2,\mathbb{Z})$ transformation. It will be interesting to study the action of $SL(2,\mathbb{Z})$ on one-parameter lift of Givental's  permutation equivariant quantum K-theory that we study in this work. We surmise that this path leads to a definition of certain {\it Quantum Elliptic Cohomology}.
\section{Conclusions and future directions}

We have presented a two-parameter deformation of the 2d GLSM/quantum cohomology correspondence. This deformation corresponds to  one-parameter refinement of permutation-equivariant quantum K-theory invariants of Givental. We have checked that the generating function of these refined invariants reduces to the generating function of unrefined invariants in $q\to 0$ limit. In doing so, we have proposed a conjectural generalization of the reconstruction theorem of Givental. We have given an expression for a 4d difference operator corresponding to 4d gauge theory and have shown that it reduces to a deformation of 3d difference equation of permutation-equivariant quantum K-theory.

In this work we report on the first of a series of explorations regarding 4d generalizations of the 2d GLSM framework. Below we list some of the next steps: 
\begin{itemize}

\item Inspired by physical considerations, we initiate the study of a two-parameter refinement of quantum cohomology which can be described by $t=\sum_{m,n\in\mathbb{Z}}t_{m,n}\tilde{P}^mq^n$ which is a one-parameter generalization of Givental's permutation-equivariant quantum K-theory.  It is important to determine the conceptual home of this expression, as it is central in defining the 4d invariants of a K\"ahler manifold $X$.

\item It is important to provide a proof of the proposition (\ref{eq:proposition}). To attempt the proof, one will need to modify the definition of quantum K-theory which incorporates the generalization of the input $t$ to $t=\sum_{m,n\in\mathbb{Z}}t_{m,n}\tilde{P}^mq^n$. The next step is to work out the relation between correlators \cite{Lee2001QuantumI} $<\tau_{k_1}(\gamma_1)...\tau_{k_n}(\gamma_n)\tau_{0}(1)>$,$<\tau_{k_1}(\gamma_1)...\tau_{k_n}(\gamma_n)>$, called the string equation. Another important relations is between the correlators $<\tau_{k_1}(\gamma_1)...\tau_{k_n}(\gamma_n)\tau_{0}(e_0)>$,$<\tau_{k_1}(\gamma_1)...\tau_{k_n}(\gamma_n)>$, called the dilaton equation, where $e_0$ is the K-class of the structure sheaf of the manifold $X$. 

\item It will be interesting to find expressions for the 4d refined invariants which are in closed form in variables $\tilde{P}$ and $q$.

\item In this manuscript, we have started exploring the role of the modular parameter, $\tau$, of $\mathbb{T}^2$. We plan to more systematically investigate its implications for the topological theory described. We anticipate that it might lead to a physical realization of a version of {\it Quantum Elliptic Cohomology} as a generalization of the ideas ellaborated in \cite{landweber1986elliptic}.

\end{itemize}

LPZ is partially supported by the U.S. Department of Energy under grant DESC0007859.


\appendix

\bibliographystyle{JHEP}
\bibliography{mybibliography}
\end{document}